\documentclass[
	paper=a4,
	DIV=12,
	abstract=on
	]{scrartcl}
\setkomafont{caption}{\small}
\setkomafont{captionlabel}{\bfseries}
\setcapindent{0pt}
\usepackage{microtype}
\usepackage{fix-cm}
\usepackage{csquotes}

\usepackage[
backend=biber,
style=numeric-comp,
sorting=none,
giveninits=true,
maxbibnames=99,
hyperref=true,
doi=true,
isbn=false,
url=false,
date=year,
]{biblatex}
\AtEveryBibitem{%
    \ifentrytype{software}{%
        \toggletrue{bbx:url}%
    }{}%
}

\usepackage{graphicx}
\input{insbox}

\usepackage{amsmath}
\usepackage{amssymb}
\usepackage[version = 4]{mhchem}

\usepackage{siunitx}
\usepackage[hidelinks]{hyperref}

\usepackage{orcidlink} 

\usepackage[
english,
sort&compress,
capitalize,
noabbrev,
nameinlink,
]{cleveref} 

\usepackage{authblk}

\author[1]{\orcidlink{0009-0003-2335-4282}\,Christopher G. O. Weiß*\textsuperscript{,}}
\author[1]{\orcidlink{0009-0002-5639-8989}\,Ellen Brennfleck}
\author[1,2]{\orcidlink{0000-0002-6540-6218}\,Bert Lägel}
\author[1]{\orcidlink{0000-0002-0725-6696}\,Laurenz Rettig}
\author[3]{\newline\orcidlink{0000-0001-8439-434X}\,Benjamin Stadtmüller}
\author[1]{\orcidlink{0000-0003-3413-5029}\,Martin Aeschlimann}
\author[3]{\orcidlink{0000-0002-7362-6448}\,Tobias Eul}

\affil[1]{Department of Physics and Research Center OPTIMAS, RPTU University Kaiserslautern-Landau, 67663 Kaiserslautern, Germany}
\affil[2]{Nano Structuring Center (NSC), RPTU University Kaiserslautern-Landau, 67663 Kaiserslautern, Germany}
\affil[3]{Experimentalphysik II, Institute for Physics, University of Augsburg, 86135 Augsburg, Germany}

\title{Reflective Metastructure Q-plate for\\Ultrashort Laser Pulses}

\date{\normalsize *Email: \href{mailto:weissc@rptu.de}{weissc@rptu.de}}

\usepackage{color}
\usepackage[normalem]{ulem}

\begin{document}
\maketitle

\begin{abstract}
\InsertBoxR{0}{%
	\hspace{0.5em}%
		\includegraphics[width=0.4\textwidth]{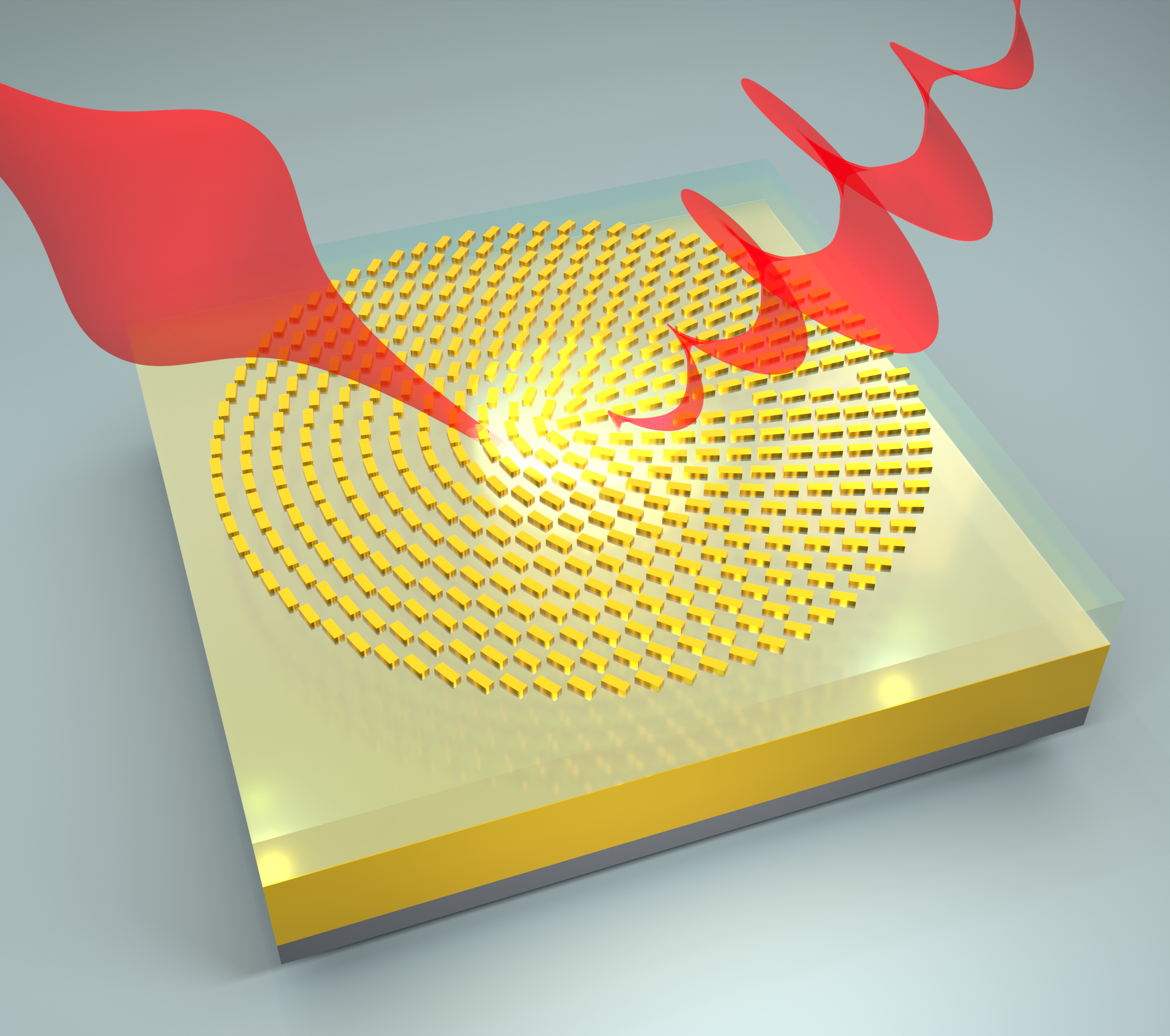}%
}
\noindent
The orbital angular momentum of light is an intriguing property for developing light driven applications.
It emerged as an independent degree of freedom by which to manipulate light and, consequently, the interaction of light with matter. Several methods exist for the generation of light carrying orbital angular momentum, mostly employing transmitting or reflecting optical components, which radially modulate the phase profile of the light. As one of such components, transmissive q-plates established themselves as standard elements due to their usability over a broad wavelength range. Here, we present our approach to building a highly reflective q-plate with a diameter of \qty{4.5}{\milli\meter} based on a plasmonic metasurface capable of converting orbital angular momentum from the nanostructure to ultrashort laser pulses without measurable temporal broadening under the investigated conditions. We highlight its working principle over a wide range of wavelengths for reflection under normal and grazing incidence.
\end{abstract}

\section*{Keywords}
orbital angular momentum, q-plate, plasmonic metasurface, reflectivity, ultrafast optics

\newpage
\twocolumn

\section{Introduction}
In optics, the angular momentum of light is commonly separated into spin angular momentum (SAM) and orbital angular momentum (OAM). SAM is associated with the polarization state of light and was first predicted by Poynting \cite{poynting1909} and experimentally verified by Beth \cite{beth1936}. For circularly polarized light, the electric field vector rotates around the propagation direction while maintaining a constant magnitude. Such states carry a spin angular momentum of \( \vec{S} = s\hbar\hat{\vec{k}} \), where \( \hat{\vec{k}} \) denotes the unit vector along the propagation direction and \( s = \pm 1 \) is the helicity of the circular polarization state.

In addition to SAM, light can possess intrinsic orbital angular momentum, which is linked to the spatial phase structure of the beam. Allen et\,al. \cite{allen1992} showed that beams with an azimuthal phase dependence of the form \( \exp{ \{\mathrm{i}\ell\phi} \} \) carry an OAM of \( \ell\hbar \) per photon, where \( \ell\in\mathbb{Z} \). In such beams, the phase forms a helical structure around the propagation axis, resulting in intertwined helical electric and magnetic field fronts. A direct consequence of this phase distribution is the emergence of a phase singularity at the beam center with vanishing intensity. Therefore, OAM beams are typically characterized by donut-like annular intensity profiles with a dark central core. The size of this singularity, and thus the diameter of the intensity ring, generally increases with increasing \( |\ell| \)\cite{padgett2004}.

The additional degree of freedom given by the OAM of light and its independence of the polarization states provides new avenues for applications involving light. The OAM can, for example, influence light-matter interaction \cite{schmiegelow2016}, give access to a multitude of entangled light states \cite{erhard2017}, increase the information depth in multiplexing \cite{wang2012a,bozinovic2013,wang2016a,weiss2026}, enable optical micromanipulation \cite{he1995,padgett2011} and facilitate high-dimensional quantum information processing \cite{erhard2017,wang2015a}. This covers many research fields, which makes it especially important to have an ample number of methods available to generate light carrying OAM suitable for the individual requirements of the respective application.

Among many possibilities, three methods established themselves for imprinting orbital angular momentum (OAM) onto light \cite{shen2019}. The most straightforward method is based on spiral phase plates (SPPs), which generate optical vortices by introducing a continuously varying azimuthal phase delay through an azimuthally increasing optical path length around the beam axis. This concept has been widely realized in transmission \cite{beijersbergen1994,sueda2004,oemrawsingh2004} and extended to reflective \cite{campbell2012,longman2020,bae2020} geometries using spiral phase mirrors. Since the imposed phase shift is wavelength dependent, conventional SPPs are typically designed for a specific center wavelength. For broadband or ultrashort pulses, spectral components farther away from the design wavelength can therefore experience reduced conversion fidelity, including effects such as topological-charge dispersion and azimuthal-angle-dependent group delay \cite{yamane2012}.

The most flexible approach is employing a liquid crystal spatial light modulator (SLM), an array of liquid crystals whose orientation can influence the amplitude, phase, and polarization of the light. The orientation is programmable for the individual crystals in the array, making it possible to imprint any phase profile onto the light \cite{konforti1988,forbes2016}. For ultrashort pulses, practical implementations are often optimized for a limited spectral range, which can reduce the conversion performance for broadband spectra \cite{efimov1995,wefers1995}. Furthermore, liquid-crystal SLMs are limited by laser-induced damage and thermally induced changes in their optical response, with the applicable power handling depending on parameters such as mirror coating, wavelength, pulse duration, repetition rate, and beam size. This constrains their direct use with ultrafast laser systems and often requires attenuation, beam expansion, or cooling \cite{beck2010}.

The third method relies on the geometric Pancharatnam-Berry phase to structure the optical field \cite{biener2002}. This phase is acquired when the polarization state of light evolves, and is therefore determined by the geometry of this evolution rather than by a conventional optical path-length difference. In Pancharatnam-Berry elements, this geometric phase is controlled locally by the orientation of an anisotropic optical element.

A prominent example is the q-plate, an optical element whose optic axis rotates with the azimuthal coordinate around a central singularity \cite{marrucci2006,marrucci2012}. The parameter \(q\) denotes the topological charge of this optic-axis pattern and specifies how many times the local optical axis rotates during one full turn around the center. This patterned birefringence locally changes the polarization state of the incident light and simultaneously imprints an azimuthally varying geometric phase. For circularly polarized input light, this results in a beam with inverted circular polarization and orbital angular momentum \(\ell = \pm 2q\), where the sign depends on the handedness of the incident polarization. In this sense, the q-plate converts spin angular momentum into orbital angular momentum and thereby couples SAM and OAM during the conversion process \cite{rubano2019}.

Because this approach is based on geometric phase rather than on a scalar optical-path delay, q-plates and related Pancharatnam-Berry elements are particularly attractive for broadband beam shaping \cite{rafayelyan2016} and for use with ultrashort laser pulses. In transmissive liquid-crystal implementations, however, the conversion efficiency still depends on the wavelength-dependent retardance \cite{sanchez-lopez2018}, and transmission through the material introduces dispersion \cite{jullien2016}. Reflective implementations of the same geometric phase concept are therefore especially appealing for ultrashort-pulse applications, as they can enable broadband operation while reducing dispersion effects in the optical path.

In this work, we propose and demonstrate an approach to creating twisted light with a large bandwidth required for ultrashort laser pulses (less than \qty{30}{\femto\second}) using a q-plate based on a reflective plasmonic metamaterial. Such a metastructure typically consists of subwavelength-sized elements that locally tailor the phase, amplitude, or polarization of impinging waves, thereby producing a macroscopic modification of the overall beam profile \cite{yu2011,aieta2012,kats2012,chen2016}. Among these, geometric-phase metasurfaces have proven to be a versatile platform for OAM generation and detection \cite{chen2018}. In the visible spectral range, such concepts have often been implemented using plasmonic nanostructures because the dimensions of the individual elements strongly depend on the target wavelength range \cite{pors2013a,zhao2011}. These offer a high degree of flexibility, as their optical response can be readily tuned through their geometric shape, while birefringent behavior emerges naturally from asymmetric designs such as nanorods \cite{jiang2014,yue2016,yue2017,liu2017}, L-shapes \cite{sung2008,karimi2014,bouchon2015,wang2015}, or elliptical nanoantennas \cite{elliott2004,wang2012,arbabi2015,walmsness2019}.

Gold nanorods are particularly well suited for this purpose, as their elongated geometry gives rise to an anisotropic plasmonic response with distinct resonances along the long and short axis. As a result, they act as plasmonic nanoantennas whose amplitude and phase response depend strongly on the polarization of the incident light \cite{link1999,huang2007,olson2015}. In ordered arrays on a substrate surface, this form-induced anisotropy leads to an effective birefringent behavior, such that the nanorods can locally control polarization conversion and phase \cite{zhao2011,zhao2013,grady2013}. If the response along the two principal axes is properly adjusted, the individual elements can behave similarly to local half- or quarter-wave plates \cite{zhao2013,pors2013b,pors2013,ding2015}. In this case, the in-plane rotation of the nanorods directly determines the geometric Pancharatnam–Berry phase imposed onto the outgoing light \cite{huang2012}.

Reflective metasurfaces for OAM generation have already been reported in the microwave and optical regimes \cite{chen2016a,yu2016}. Yue et al. demonstrated reflective plasmonic metasurfaces based on oriented gold nanorods for generating vector vortex beams and superpositions of OAM states \cite{yue2016,yue2017}. Other studies have reported reflective metasurfaces for rotational Doppler measurements, vector-vortex-beam generation, and focused vortex beams \cite{georgi2017,heiden2019,yang2019,ding2020,tang2020}. Thus, re\-flec\-tion-mode OAM generation and the use of plasmonic nanorods are established concepts.

The distinguishing feature of this work is the experimental realization of a large-area reflective plasmonic q-plate that brings together high reflectivity, broadband operation, and verified compatibility with ultrashort laser pulses in a single metasurface platform.
By combining the high-reflectivity properties of the gold/\allowbreak dielectric/\allowbreak nanorods interface reported in \cite{ding2015,zheng2015a} with the intrinsic birefringence of the nanorods, we establish a reflective q-plate achieving reflectivities above \qty{70}{\percent}.
Its \qty{4.5}{\milli\meter} aperture represents a significant step towards practical implementation, enabling direct illumination with millimeter-sized laser beams without tight focusing and seamless integration into conventional free-space laser systems.

Beyond its large-area design, the metastructure provides versatile control over the reflected light, generating OAM-carrying beams with inverted helicity and $\ell=\pm1$ under circularly polarized illumination, as well as radial and azimuthal vector beams under linearly polarized illumination.
We demonstrate broadband OAM generation across \qtyrange{600}{1000}{\nano\meter} at both near-normal (\qty{5}{\degree}) and oblique (\qty{45}{\degree}) incidence, with operation extending down to \qty{550}{\nano\meter} at \qty{45}{\degree}.
Furthermore, the metastructure maintains a total reflectivity exceeding \qty{60}{\percent} for wavelengths above \qty{700}{\nano\meter}.

Crucially, we go beyond spectral characterization by directly demonstrating the temporal performance of the device using approximately \qty{26}{\femto\second} pulses at \qty{800}{\nano\meter} and an incidence angle of \qty{45}{\degree}.
Under these conditions, reflection introduces no measurable pulse broadening, in clear contrast to the measurable pulse-duration increase observed when using a commercial transmissive q-plate as a reference.

Taken together, these results establish a new platform for broadband reflective control of orbital angular momentum and vector-beam generation, overcoming key limitations associated with conventional transmissive q-plates.
The combination of millimeter-scale aperture, high reflectivity, broadband operation at practically relevant incidence angles, and experimentally verified ultrashort-pulse preservation positions our device as a promising building block for next-generation ultrafast photonic and laser systems.

\section{Design of the Metasurface}
\begin{figure*}[tb]
	\includegraphics{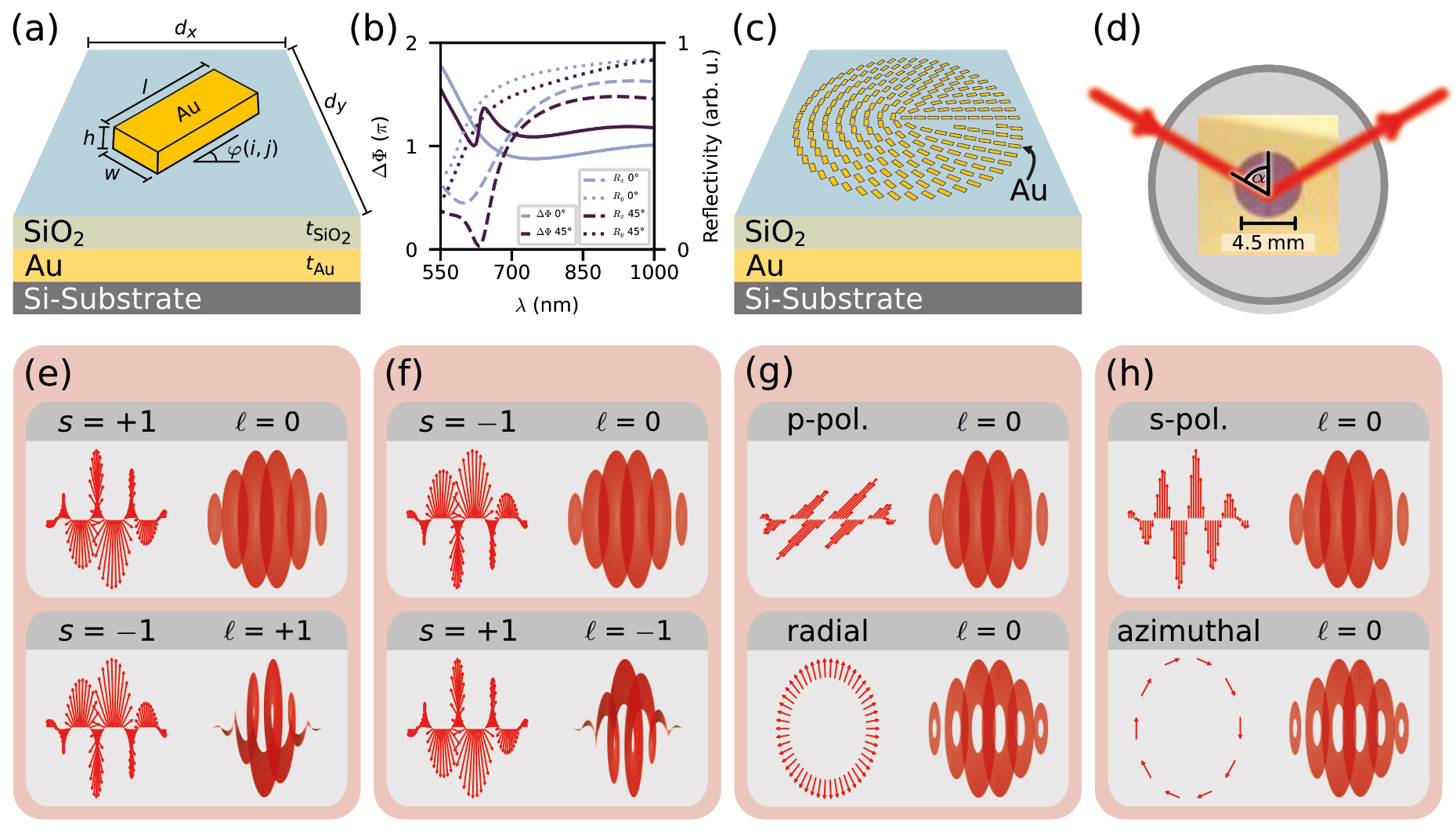}
	\caption{Operating principle of the reflective plasmonic q-plate with a diameter of \qty{4.5}{\milli\meter}.
    (a) Building block of the metastructure consisting of a single gold nanorod on a \ce{SiO2}/\ce{Au}-reflector stack on a \ce{Si}-substrate.
    (b) Simulated phase retardance $\Delta\Phi$ between the reflected fields and reflection coefficients $R_x$ and $R_y$ for incident light polarized along the long $(x)$ and short $(y)$ axes of the nanorod, respectively. Results are shown for incidence angles of \qty{0}{\degree} and \qty{45}{\degree}.
    (c) Metastructure with azimuthally oriented gold nanorods that impose a spatially varying geometric phase upon reflection.
    (d) Illustration of the metastructure glued to a mirror.
    The diameter of the structure is \qty{4.5}{\milli\meter} and the light can be reflected under an angle of incidence $\alpha\in\{\qty{5}{\degree},\qty{45}{\degree}\}$.
    (e-h)
    Top panel shows the phase front and polarization of the incident beam, whereas the bottom panels display the corresponding reflected states.
    Circularly polarized input light is converted into OAM-carrying beams with opposite helicity and \( \ell=\pm 1 \), while linearly polarized input states give rise to radial and azimuthal vector beams with zero net orbital angular momentum.}
	\label{fig:introFigure}
\end{figure*}

The operating principle of the reflective metasurface q-plate is illustrated in \cref{fig:introFigure}.
The structure is comprised of periodically arranged building blocks consisting of a gold nanorod in the center placed on a dielectric spacer above a gold back reflector on a \ce{Si}-substrate (\cref{fig:introFigure}(a)).
The block has side lengths of $d_x=d_y=\qty{250}{\nano\meter}$.
The dielectric spacer is made out of SiO$_2$ with a thickness of $t_\mathrm{SiO_2}=\qty{95}{\nano\meter}$.
The gold mirror layer beneath has a thickness of $t_\mathrm{Au}=\qty{300}{\nano\meter}$.
The nanorod has a length, width and height of $l=\qty{200}{\nano\meter}$, $w=\qty{85}{\nano\meter}$ and $h=\qty{30}{\nano\meter}$, respectively.
The parameters were initially chosen based on reference \cite{zheng2015a} but slightly altered to accommodate for our fabrication capabilities and, at the same time, optimizing the functionality of the metastructure.

The optimization process was done via finite-difference time domain simulations \cite{meep} of the optical response of the individual nanorod.
\cref{fig:introFigure}(b) shows the wavelength dependent response including the phase retardance $\Delta\Phi$ for a reflection of light polarized along the long $(x)$ and short $(y)$ axis of the nanorod.
The graph additionally displays the corresponding reflection coefficients $R_x$ and $R_y$.
Both simulations were done for the normal incidence (\qty{0}{\degree}) and grazing incidence (\qty{45}{\degree}) geometry.
The normal-incidence calculation serves as an approximation to the experimental near-normal-incidence geometry at \qty{5}{\degree}.
For both geometries, the retardance is very close to the desired value of $\pi$ for ideal half-wave plate behavior for wavelengths above \qty{600}{\nano\meter}, making it very suitable for the use in the metastructure.
For the lower wavelength range, the retardance starts to deviate from the ideal half-wave plate behavior.
Section S1 of the Supporting Information contains additional simulations for varying dimensions of the nanorod and the dielectric spacer.

\cref{fig:introFigure}(c) shows the metastructure design with an azimuthal pattern of the local in-plane orientation of the nanorods, which induces a spatially varying geometric Pancharatnam–Berry phase upon reflection.
To realize the reflective q-plate, the orientation and position of the nanorods needs to create the same azimuthal birefringence landscape as a conventional q-plate.
As for conventional q-plates, the implemented transformation is not fixed to a single output state but is entirely determined by the polarization of the incident light, such that different input states are mapped to distinct output modes by the same structure.
The radial distance between the individual nanorods corresponds to the side length \( d_x=d_y=d_r = \qty{250}{\nano\meter} \) of the nanorod building block.
The overall metasurface with a radius of \(R\) therefore consists of \( n = \frac{R}{d_r}\) circles of nanorods.
Within each of these circles the nanorods have an azimuthal distance of \( d_a^i = \frac{1}{i} \) in radians for \( i \in \{ 1,\dots{},n \} \).
Assuming \(m\) nanorods fit into the \(i\)-th circle, then the rotation angle of a nanorod amounts to
\[ \varphi(i,j) = \frac{j\cdot d_a^i}{2 \pi}\cdot \ell = \frac{j}{2i}\cdot\ell \]
for \( j \in \{ 0,\dots{},m-1 \} \) and a metasurface able to generate an OAM of the order \(\ell\).
\cref{fig:introFigure}(d) additionally illustrates how the metastructure is used as an optical component glued with carbon tape on a standard mirror.
The diameter of our metastructure is \qty{4.5}{\milli\meter} and we can illuminate it under different angles of incidences $\alpha$.

The panels in \cref{fig:introFigure}(e-h) compare the spin and orbital angular momentum in terms of their phase profile and polarization state before (top half of the image) and after reflection (bottom half of the image).
Circularly polarized input light with zero orbital angular momentum is converted into a beam with opposite spin angular momentum and a helical phase front with \( \ell=\pm 1 \), where the sign depends on the handedness of the incident polarization.
In contrast, linearly polarized input states are transformed into vector vortex beams, such as radially and azimuthally polarized modes. In the ideal case, these output states comprise equal-weight superpositions of opposite circular polarizations carrying opposite orbital angular momenta, resulting in zero net orbital angular momentum.
In this way, the metasurface couples the polarization state of the incident light to the phase structure of the reflected beam.

\begin{figure}[t]
	\includegraphics{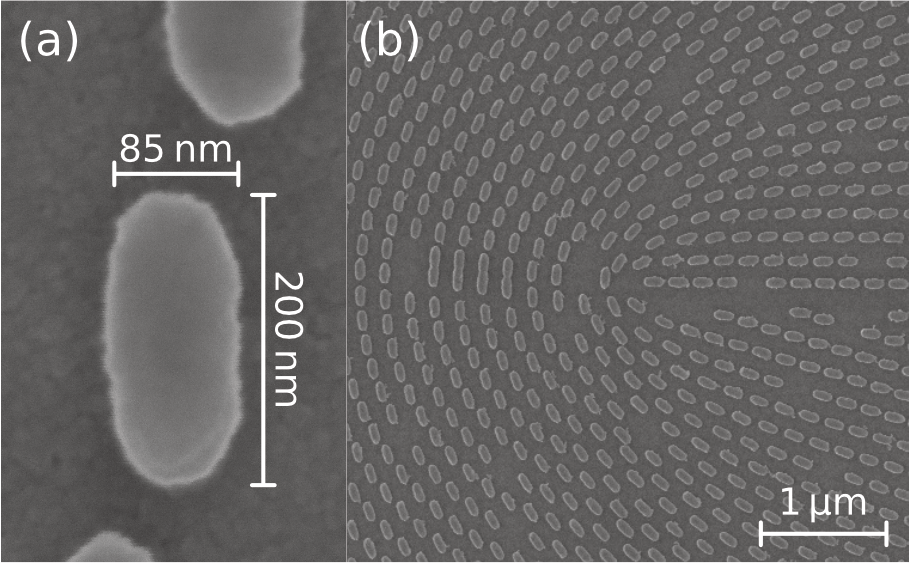}
	\caption{Scanning electron microscopy (SEM) images of the fabricated reflective plasmonic metasurface q-plate based on arrays of gold nanorods on a \ce{SiO2}/\ce{Au}-substrate. (a) High-magnification SEM image of individual gold nanorods, indicating their typical dimensions (\( l = \qty{200}{\nano\meter} \), \( w = \qty{85}{\nano\meter} \)). (b) SEM overview of the central region of the metasurface showing the azimuthally varying nanorod orientations that implement the q-plate phase profile.}
	\label{fig:SEMoverview}
\end{figure}

The fabrication was carried out at the Nanostructuring Center (NSC) of the RPTU Kaiserslautern–Landau.
First, a gold mirror was deposited onto a \qtyproduct{1x1}{\centi\metre\squared} p-doped \ce{Si}-wafer by magnetron sputtering.
To promote adhesion between the substrate and the gold film, a \qty{5}{\nano\meter} chromium layer was sputtered beforehand.
Subsequently, a \ce{SiO2}-spacer layer was deposited using the same magnetron sputtering process.
For precise thickness control of the \ce{SiO2}-layer, the deposition rate of the sputtering process was determined beforehand.
The sample was then spin-coated with a positive electron-beam resist (PMMA \textit{AR-P 679.04}, 950k dissolved at \qty{4}{\percent} in ethyl lactate; \textit{Allresist GmbH}).
The individual nanorods forming the metasurface pattern were defined in an electron-beam lithography (EBL) step.
Due to the large diameter of the optical element (\qty{4.5}{\milli\meter}), the structure was divided into \qtyproduct{200 x 200}{\micro\meter\squared} write fields, which were stitched together during exposure.
Writing approximately \num{250} million nanorods required about \qty{27}{\hour}.
The exposures were performed using a \textit{RAITH Voyager} system.

After development, an \ce{O2} plasma cleaning step was applied to remove remaining resist residues.
The exposed regions were then deposited by electron-beam evaporation with a \qty{30}{\nano\meter} gold film.
Lift-off was carried out in acetone, followed by rinsing with isopropanol and blow drying under flow of \ce{N2}. 
\cref{fig:SEMoverview} shows two representative scanning electron microscopy (SEM) images of the fabricated metasurface.
\cref{fig:SEMoverview}(a) displays a single nanorod together with its geometric dimensions.
\cref{fig:SEMoverview}(b) shows the central region of the q-plate, highlighting the local rod orientations determined by the radial coordinate.
A small number of nanorods are missing, presumably due to non-optimal process conditions during the electron-beam lithography process.
Section S2 of the Supporting Information further discusses the influence of fabrication issues on the performance of the metastructure.

\section{Experimental Setup}\label{sec:ExperimentalSetup}
To characterize the optical response of the fabricated reflective q-plate, wavelength- and polarization-dependent measurements were performed.
For this purpose, a femtosecond laser system covering a broad spectral range was used, allowing us to investigate the spatial intensity distribution, reflectivity, and polarization transformation of the reflected beam.

\begin{figure}[t]
	\includegraphics{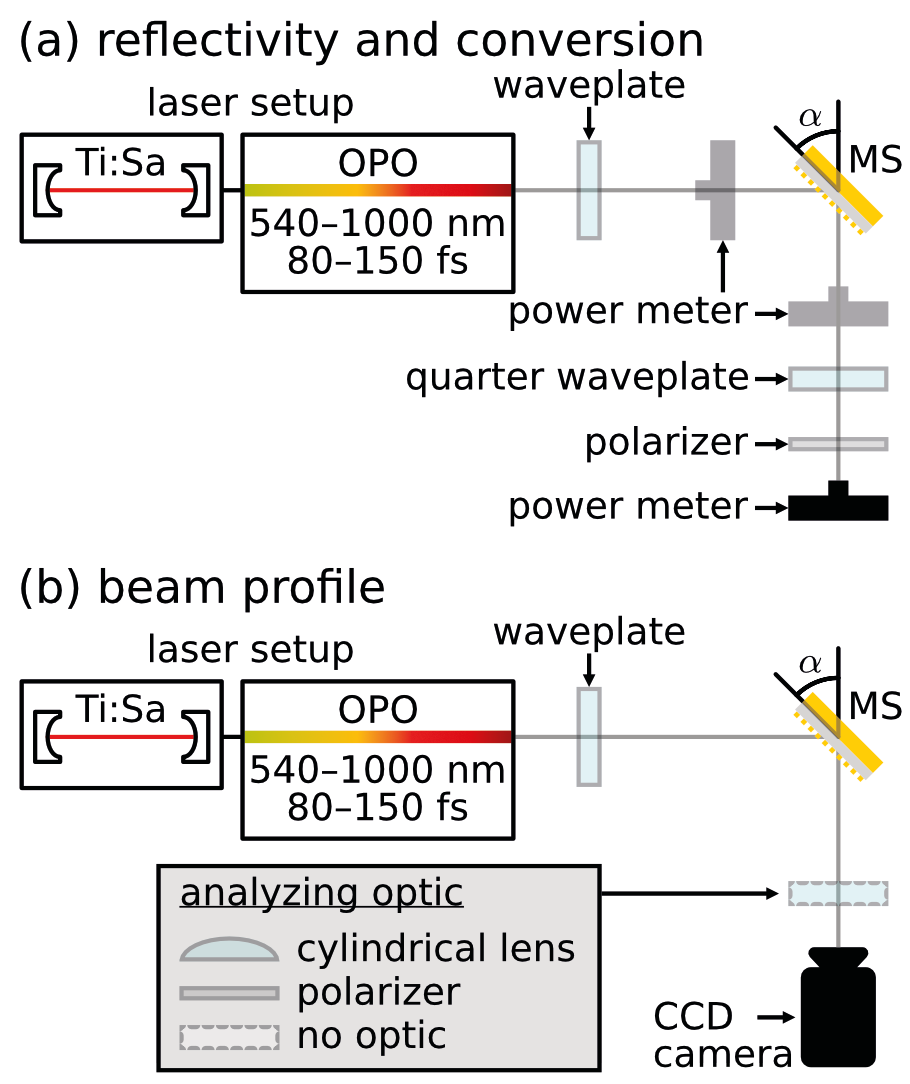}
	\caption{Schematic of the optical setup for characterizing the reflective q-plate. A tunable \ce{Ti}:\ce{Sa}-oscillator with an OPO provides femtosecond pulses from \qtyrange{540}{1000}{\nano\meter} with pulse durations of \qtyrange{80}{150}{\femto\second}. The incident polarization is controlled by a quarter- or half-wave plate, and the metasurface (MS) is illuminated at \(\alpha = \qty{45}{\degree}\) or \(\qty{5}{\degree}\). (a) Power-meter setup for reflectivity and cross-polarization measurements, with a quarter-wave plate and polarizer used for the latter. (b) Beam-profile setup with CCD-camera detection. For further analysis, either no additional optic, a polarizer, or a cylindrical lens can be placed in front of the camera.}
	\label{fig:setup}
\end{figure}

\cref{fig:setup} shows the experimental setup used for the optical characterization of the reflective q-plate.
A tunable titanium-sapphire (\ce{Ti}:\ce{Sa}) oscillator (\textit{Spectra Physics}) was used in combination with an optical parametric oscillator (OPO, \textit{Radiantis}).
Together, these systems provide femtosecond pulses in a wavelength range from \qtyrange{540}{1000}{\nano\metre} with pulse durations between \qtyrange{80}{150}{\femto\second}, which is well suited to investigate the broadband response of the metasurface.

\cref{fig:setup}(a) depicts the setup used for the reflectivity and polarization-conversion measurements.
After passing the laser source and OPO, the incident polarization is adjusted by either a half-wave plate (linear pol.) or a quarter-wave plate (circularly pol. ) before the beam is directed onto the metasurface sample under an angle of incidence \(\alpha\).
For the reflectivity measurements, the reflected power is recorded directly with a power meter (\textit{Ophir Vega}). For the cross-polarization measurements, a quarter-wave plate and a polarizer are inserted into the reflected beam path before the power meter to determine the wavelength-dependent conversion of the q-plate.

For the cross-polarization measurements, the first wave plate is a quarter-wave plate that converts the initially linearly polarized light into circularly polarized light. The metasurface converts part of the incident light into the orthogonal circular polarization state while imparting orbital angular momentum to this component.

The remaining light retains its original polarization state. A second quarter-wave plate placed after the metasurface converts the two circular polarization components into orthogonal linear polarization states, which are selected successively using a polarizer and measured with a power meter.

\begin{figure*}[t]
	\includegraphics{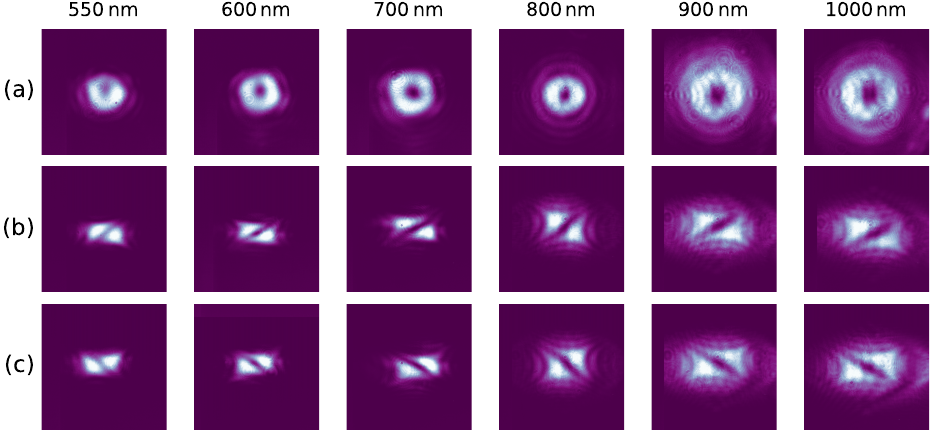}
	\caption{Wavelength dependent OAM generation under \qty{45}{\degree} illumination. (a) Beam profiles after reflection of \(\sigma^+\)-light (\(s=+1\)) on the q-plate. (b) Beam profile from (a) with inserted cylindrical lens. (c) Beam profile after reflection of \(\sigma^-\)-light (\(s=-1\)) with inserted cylindrical lens. Each beam-profile image was individually normalized to its respective maximum intensity.}
	\label{fig:beamProfiles45}
\end{figure*}

The conversion efficiency \( \eta \) is calculated as
\[
\eta=\frac{P_{\mathrm{cross}}}{P_{\mathrm{inc}}},
\]
where \(P_{\mathrm{inc}}\) is the incident power and \(P_{\mathrm{cross}}\) is the power in the converted polarization component. In addition, the polarization conversion ratio $PCR$ is determined as
\[
PCR=\frac{P_{\mathrm{cross}}}{P_{\mathrm{cross}}+P_{\mathrm{co}}},
\]
where \(P_{\mathrm{co}}\) denotes the power remaining in the unconverted, co-polarized component.

\cref{fig:setup}(b) shows the setup used for beam profile measurements.
In this configuration, the reflected light is sent to a CCD camera to record the spatial intensity distribution.
As in the reflectivity measurements, the incident polarization is controlled by either a half-wave plate or a quarter-wave plate, and the angle can be set to the two values described above.
Depending on the measurement, additional analyzing optics can be inserted in front of the camera.
A cylindrical lens is used to analyze the sign and magnitude of the orbital angular momentum, while a polarizer is used to investigate the polarization state of the reflected beam. Measurements can also be performed without any additional optic to directly record the beam profile.

To investigate possible pulse broadening by the q-plate, we additionally used a separate Titanium-Sapphire Oscillator (\textit{Spectra Physics}) providing shorter pulses of approximately \qty{26}{\femto\second} at a center wavelength of \qty{800}{\nano\meter}. The autocorrelation measurements were performed at an incidence angle of \qty{45}{\degree} using an \textit{APE} \textit{PulseCheck} autocorrelator.

\section{Results and Discussion}
In typical optical assemblies, reflective components are commonly employed either at near-normal incidence (\qty{5}{\degree}-reflection) or at an incidence angle of \qty{45}{\degree}, corresponding to a \qty{90}{\degree}-beam deflection.
These configurations are widely used as they largely preserve the polarization state of the incident light upon reflection and therefore provide a reliable basis for polarization-sensitive applications.
For this reason, we restrict our investigation to these two representative reflection geometries, which reflect common experimental practice.
Within these configurations, a variety of vortex beams can be generated by the plasmonic metastructure depending on the input polarization, as shown in \cref{fig:introFigure}.

\begin{figure}[t]
	\includegraphics{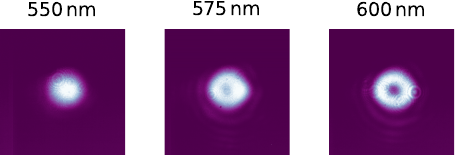}
	\caption{Wavelength dependent OAM generation under \qty{5}{\degree} illumination. Beam profile after reflection of \(\sigma^+\)-light (\(s=+1\)) on the q-plate for the wavelengths \qtylist{550;575;600}{\nano\metre} from left to right. Each beam-profile image was individually normalized to its respective maximum intensity.}
	\label{fig:beamProfilesNI}
\end{figure}

\begin{figure*}[t]
	\includegraphics{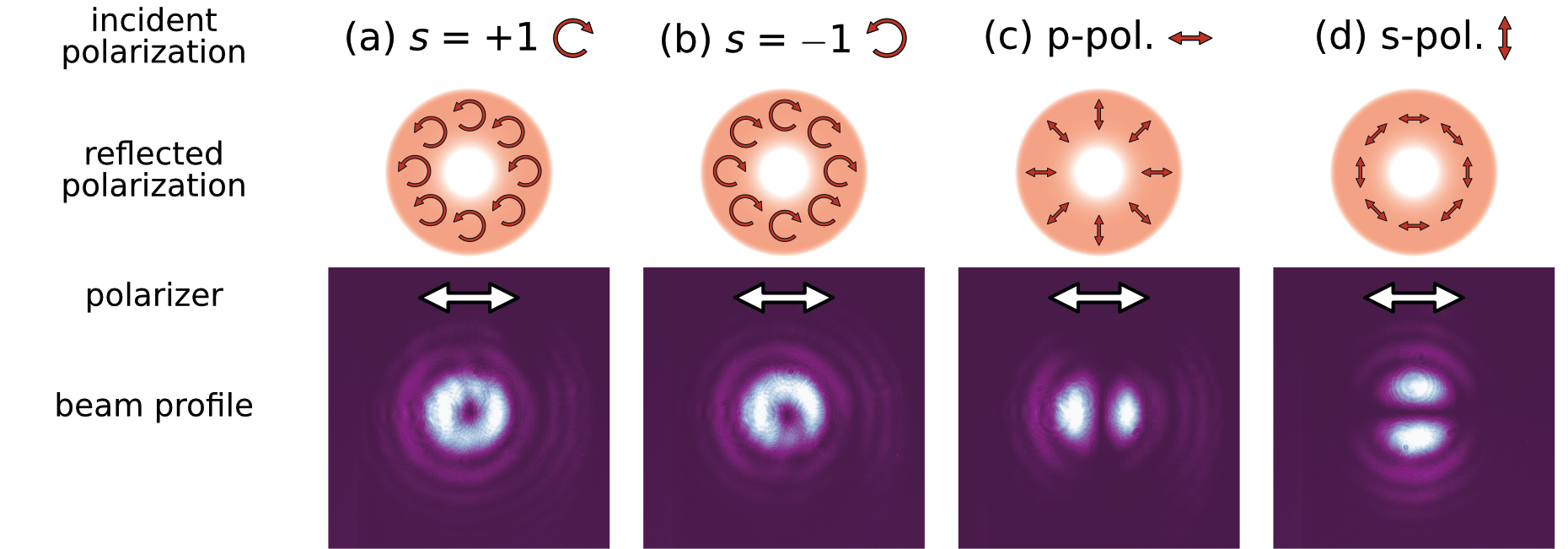}
	\caption{Polarization-dependent beam profiles recorded at \qty{800}{\nano\meter} for an angle of incidence of \qty{45}{\degree}. From top to bottom, each column shows the incident polarization, the expected polarization after reflection from the q-plate, the orientation of the analyzing polarizer, and the measured beam profile. Circularly polarized excitation with (a) \(s=+1\) and (b) \(s=-1\) is converted into the opposite circular polarization state, while the donut-shaped beam profile is preserved. For (c) p-polarized excitation, the reflected beam is radially polarized, resulting in transmission of the left and right parts of the beam. For (d) s-polarized excitation, the reflected beam is azimuthally polarized, leading to transmission of the upper and lower parts. Each experimentally recorded beam-profile image was individually normalized to its respective maximum intensity.}
	\label{fig:beamProfilesPolarization}
\end{figure*}

First, we will focus on the generation of vortex beams with orbital angular momentum (OAM).
Hence, we analyze the performance of the q-plate under circularly polarized light illumination.
\cref{fig:beamProfiles45} shows exemplary data from our available wavelength range for a reflection of the light under \qty{45}{\degree} grazing incidence.
All experimentally recorded beam-profile images presented in \cref{fig:beamProfiles45,fig:beamProfilesNI,fig:beamProfilesPolarization} were individually normalized to the maximum intensity of the respective beam profile because the available laser output power varies with wavelength.
As a result of this individual normalization, the displayed brightness cannot be quantitatively compared between different panels.
Separately, isolated hot pixels were removed prior to normalization to prevent detector artifacts from determining the normalization maximum.
The top row (a) contains the experimentally obtained beam profiles for \(\sigma^+\)-polarized (\(s=+1\)) incoming light.
All beam profiles exhibit the characteristic donut mode for light carrying orbital angular momentum.
For the smallest wavelength of \qty{550}{\nano\meter} the beam profile seems to be a superposition of a donut with a background signal, possibly from a non perfect conversion of the light at the nanostructures.
The conversion efficiency will be further discussed below.
Apart from the donut shape, all profiles additionally show multiple rings with increasing radius associated with higher orders of the radial index in the Laguerre-Gaussian modes.
The increase of the singularity and overall beam size mirrors the wavelength dependent beam properties of the light source.
It is remarkable that the q-plate maintains its functionality across the entire spectral range from \qtyrange{550}{1000}{\nano\meter}, thereby demonstrating the broadband capability of the plasmonic metasurface.
This behavior highlights a key advantage of reflective plasmonic metasurfaces, which are particularly well suited for applications involving broadband or ultrashort optical pulses.

To confirm that the light indeed carries orbital angular momentum, we additionally monitored the beam profiles after inserting the cylindrical lens, a well-established method to probe orbital angular momentum of light beams \cite{denisenko2009}. 
\cref{fig:beamProfiles45}(b) and (c) show these beam profiles for \(\sigma^+\)- and \(\sigma^-\)-polarized light (\(s=\pm1)\), respectively.
The tilted, triangular-shaped beam profiles without any intermediate stripes represent the characteristic signature of light carrying orbital angular momentum.
The orientation of the triangles further confirms the inversion of the OAM sign upon switching the handedness of the circular polarization.
Altogether, these results demonstrate that the presented metasurface enables robust spin-to-orbital angular momentum conversion over a broad spectral range upon reflection, making it particularly suitable for broadband and ultrafast photonic applications.

We now extend our analysis to the second experimental configuration, namely reflection of the beam close to normal incidence.
In practice, a small finite angle is required in the experimental implementation to separate incident and reflected beams. 
Therefore, we approximate the normal incident case by measurements at an incidence angle of \(\qty{5}{\degree}\), which can be regarded as near-normal incidence.

The corresponding beam profiles for this near-normal incidence configuration are presented in \cref{fig:beamProfilesNI}.
In this case, the q-plate does not generate OAM for the reflection of light with \qty{550}{\nano\meter} and only a partial generation is possible for \qty{575}{\nano\meter}.
For all investigated wavelengths from \qtyrange{600}{1000}{\nano\meter}, the reflected beam exhibits the characteristic donut-shaped mode.
The reason for this reduction in broadband functionality lies in the wavelength dependent phase retardance of the individual nanorod (see \cref{fig:introFigure}(b)).
For normal incidence the nanorod already diverges from ideal half-wave plate behavior for larger wavelengths than for the grazing incidence geometry.

\begin{figure}[t]
\centering
	\includegraphics{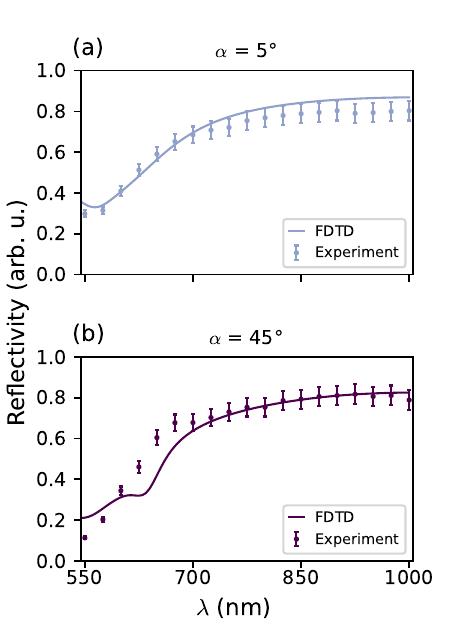}
	\caption{Measured and simulated wavelength-dependent reflectivity $R$ of the metasurface q-plate. (a) Near-normal-incidence configuration: experimental data acquired at \qty{5}{\degree} and simulation performed at \qty{0}{\degree}. (b) Grazing-incidence configuration: experimental data and simulation at \qty{45}{\degree}.}
	\label{fig:reflectivity}
\end{figure}

After confirming the functionality of the q-plate, we characterized its efficiency in terms of its wavelength dependent reflectivity followed by an analysis of the efficiency of conversion to the $\ell=1$ mode.
\cref{fig:reflectivity}(a) and (b) show the measured reflectivity for the two cases of near-normal and grazing incidence, respectively.
In both cases, the q-plate reflects more than \qty{60}{\percent} for wavelengths larger than \qty{700}{\nano\meter} with a fast drop towards lower wavelengths.

The experimental data are overlaid with the simulated reflectivity calculated from the reflection coefficients $R_x$ and $R_y$ shown in \cref{fig:introFigure}(b).
The simulations at \qty{0}{\degree} and \qty{45}{\degree} are compared with measurements at \qty{5}{\degree} and \qty{45}{\degree}, respectively.
Taking into account all rotation angles $\varphi$ of all $n$ nanorods in the metastructure, the total reflectivity is given by 
\begin{align*}
    R = \frac{1}{n} \sum_{i=1}^n \left( R_x\cos^2{\varphi_i} + R_y\sin^2{\varphi_i}\right).
\end{align*}

Both the experimental and simulated reflectivity mimics the reflectance of pure gold, albeit with an overall lower maximum.
We assume this discrepancy in comparison to pure gold stems from scattering at the nanorod upon backreflection from the mirror layer and since the measured reflectivity fits quite well to the simulations, especially for wavelengths above \qty{700}{\nano\meter}, we rule out a large influence of the sample fabrication on the reflectivity of the structure (see section S2.3 of the Supporting Information for an extended discussion of potential fabrication issues).
The reflectivity above \qty{700}{\nano\meter} is also relatively high compared to other plasmonic metasurfaces \cite{karimi2014,zhang2019b,ebel2023}.
For lower wavelengths it might be largely improved by replacing gold with materials such as silver or even aluminum.
The latter might prove especially useful for designing q-plates in the ultraviolet regime, although surface oxidation would have to be carefully considered.

\begin{figure*}[t]
    \includegraphics{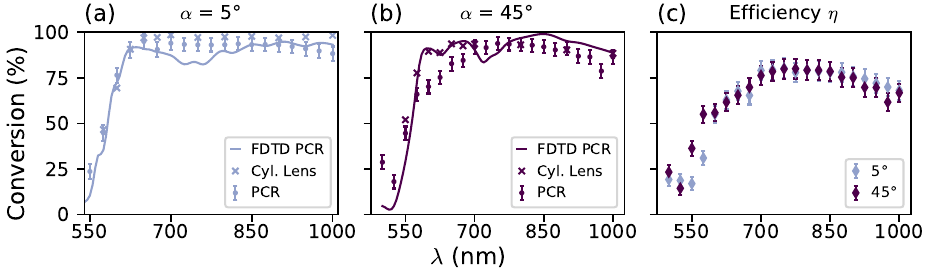}
    \caption{Conversion from the incoming trivial Gauss mode to the $\ell=1$ mode upon reflection at the metasurface.
    (a) Near-normal incidence geometry.
    (b) Grazing incidence geometry.
    Circles and lines show the polarization conversion ratio obtained from using the setup in \cref{fig:setup}(a) and FDTD simulations, respectively.
    Crosses indicate conversion ratios extracted from complementary cylindrical-lens measurements acquired for the beam profiles shown in \cref{fig:beamProfiles45} and \cref{fig:beamProfilesNI}.
    (d) Efficiency $\eta$ for both angles of incidence.
    }
    \label{fig:conversion}
\end{figure*}

The second useful quantity to discuss the device performance is the polarization conversion ratio $PCR$.
Since the helicity of the beam is expected to change its sign upon reflection, we can estimate the conversion ratio to the $\ell=1$ mode based on the amount of $\sigma^+$ and $\sigma^-$ light in the reflected beam.
For example, for $\sigma^+$-polarized input light, the converted $\sigma^-$ component is expected to dominate over the unconverted $\sigma^+$ component.
We measure the power in each component using a quarter-wave plate and a polarizer (see \cref{fig:setup}(a)) and calculate the polarization conversion ratio according to the definition in \cref{sec:ExperimentalSetup}.
The $PCR$ values (circles) in \cref{fig:conversion}(a-b) are in the range of \qtyrange{80}{90}{\percent} for the wavelength range of the \ce{Ti}:\ce{Sa} oscillator system \qtyrange{650}{1000}{\nano\meter} and both incidence angles with a higher ratio for normal incidence.
Therefore, our device converts most of the beam to the desired $\ell=1$ output mode across this wavelength range.

Further FDTD simulations \cite{Tidy3d} of the $PCR$ property reproduce this general trend with values above \qty{80}{\percent} in the same wavelength range.

Additionally, we made an effort to estimate the conversion ratio from complementary cyl\-in\-dri\-cal-lens measurements acquired for the beam profiles shown in \cref{fig:beamProfiles45} and \cref{fig:beamProfilesNI}. 
Section S3 of the Supporting Information contains a description of the extraction method of the conversion ratios from line profiles through these beam patterns.
\cref{fig:conversion}(a-b) displays the extracted conversion ratios as crosses and we can immediately see that they reproduce the helicity ratios, validating our approach.
For wavelengths lower than \qty{600}{\nano\meter}, the beam images already exhibit less pronounced donut modes hinting at incomplete conversion to the $\ell=1$ mode, especially in the normal incidence geometry.
The extracted conversion ratios support this claim with values dropping down to \qty{45}{\percent} at \qty{575}{\nano\meter}.

Finally, we extracted the device efficiency $\eta$ (\cref{fig:conversion}(c)) as described in section \ref{sec:ExperimentalSetup}, whose value is influenced by both the reflectivity and the polarization conversion ratio.
For both geometries, the efficiency exceeds \qty{60}{\percent} for wavelengths above \qty{650}{\nano\meter}.
Especially, the range between \qtyrange{700}{850}{\nano\meter} shows a stable efficiency of approximately \qty{80}{\percent}.
For wavelengths below \qty{600}{\nano\meter} on the other hand, the efficiency starts to decrease significantly with slightly higher values for the grazing incidence geometry.
Nevertheless, we can see that the metasurface operates quite efficiently over a broad wavelength range.

This broadband capability of our plasmonic q-plate suggest that it may be a suitable optic for the reflection of ultrashort laser pulses.
Conventional q-plates, on the other hand, typically lengthen the pulse duration upon transmission of the pulses.
To investigate possible pulse broadening, intensity autocorrelation measurements were performed at a center wavelength of \qty{800}{\nano\meter} and an incidence angle of \qty{45}{\degree}. The autocorrelation FWHM was \qty{37.1}{\femto\second} for the reference pulse, \qty{37.1}{\femto\second} after reflection from the plasmonic metasurface q-plate, and \qty{42.4}{\femto\second} after transmission through a commercial q-plate (\textit{Thorlabs}, \textit{WPV10L-780}). Assuming a Gaussian pulse profile, these values correspond to pulse durations of approximately \qty{26.2}{\femto\second}, \qty{26.2}{\femto\second}, and \qty{30.0}{\femto\second}, respectively. Thus, under the investigated conditions, no measurable increase in pulse duration was observed upon reflection from the plasmonic metasurface q-plate. Although the increase of approximately \qty{4}{\femto\second} introduced by the transmissive q-plate appears modest, dispersion-induced broadening generally becomes increasingly relevant for shorter input pulses. The absence of measurable broadening upon reflection under the investigated conditions therefore highlights the potential of the reflective q-plate for applications involving pulse durations below \qty{20}{\femto\second}. The corresponding autocorrelation traces are provided in Section S4 of the Supporting Information.

So far, we have focused the discussion about the properties of our plasmonic q-plate on the reflection of circularly polarized light since we are interested in generating light with OAM.
Q-plates, however, possess additional functionalities when illuminating them with linearly polarized light (see \cref{fig:introFigure}). In this case, the converted light still exhibits the typical donut shaped beam profile but, rather than changing its orbital angular momentum, the q-plate alters the polarization state of the incoming light. Depending on the orientation of the linear polarization with respect to the metastructure, the interaction with the q-plate results in either radially or azimuthally polarized light.

To investigate this behavior, we replaced the cylindrical lens in front of the CCD camera with a polarizer set to transmission of p-polarized light. \cref{fig:beamProfilesPolarization} shows the polarization-dependent beam profiles recorded at \qty{800}{\nano\meter} for an angle of incidence of \qty{45}{\degree}. Each column is arranged from top to bottom as follows: the incident polarization, the expected polarization distribution after reflection from the q-plate, the orientation of the analyzing polarizer, and the experimentally recorded beam profile. For circularly polarized excitation, shown in \cref{fig:beamProfilesPolarization}(a) for \(s=+1\) and in \cref{fig:beamProfilesPolarization}(b) for \(s=-1\), the reflected beam is circularly polarized with reversed helicity and retains its donut-shaped intensity profile.
For \(\sigma^+\)-polarized excitation (\(s=+1\)), the reflected beam has \(\sigma^-\) polarization. The analyzing polarizer reduces the transmitted intensity but preserves the donut-shaped intensity profile, as shown in \cref{fig:beamProfilesPolarization}(a).
An analogous behavior is observed for \(\sigma^-\)-polarized excitation (\(s=-1\)) in \cref{fig:beamProfilesPolarization}(b). For p-polarized excitation in \cref{fig:beamProfilesPolarization}(c), the reflected beam is radially polarized, so that a horizontally oriented polarizer mainly transmits the left and right parts of the beam. For s-polarized excitation in \cref{fig:beamProfilesPolarization}(d), the reflected beam is azimuthally polarized, and the same polarizer therefore mainly transmits the upper and lower parts of the beam.

\section{Conclusion}
Our plasmonic metasurface represents a new approach for the generation of light carrying orbital angular momentum (OAM) over a broad wavelength range upon reflection. This surface consists of a three-layer system comprised of a gold and a silicon dioxide film as well as a top layer of gold nanorods, which individually behave like a half-wave plate upon rotation. The arrangement of the rods on the surface regarding this rotation mimics the radially distributed fast axis profile of a typical q-plate.

The fabricated q-plate has a diameter of \qty{4.5}{\milli\meter}, allowing direct illumination with mil\-li\-me\-ter-sized laser beams and facilitating its integration into conventional free-space optical setups. OAM generation with \( \ell = \pm 1 \) was demonstrated from \qtyrange{600}{1000}{\nano\meter} at near-normal incidence (\qty{5}{\degree}) and from \qtyrange{550}{1000}{\nano\meter} at \qty{45}{\degree}, with incomplete conversion at the shortest wavelength. The sign of the generated OAM is controlled by the handedness of the incident circular polarization.
At the same time, it mimics the wavelength dependent reflectance of gold at least qualitatively with the reflectance maximum being about \qty{25}{\percent} lower.
The polarization conversion ratio reaches approximately \qtyrange{80}{90}{\percent} between \qtyrange[range-phrase={ and }]{700}{900}{\nano\meter} for both incidence angles, consistent with estimates obtained from the cyl\-in\-dri\-cal-lens beam profiles.
The device efficiency resulting from both the reflectivity and the polarization conversion ratio is also in the order of \qty{80}{\percent} across a large wavelength range between \qtyrange[range-phrase={ and }]{700}{850}{\nano\meter} and it exceeds \qty{60}{\percent} for all wavelengths above \qty{650}{\nano\meter}.
At a center wavelength of \qty{800}{\nano\metre} and an incidence angle of \qty{45}{\degree}, autocorrelation measurements yield the same pulse duration of approximately \qty{26.2}{\femto\second} before and after reflection from the metasurface. In comparison, transmission through the commercial q-plate tested here leads to measurable pulse broadening. These measurements demonstrate compatibility with ultrashort pulses without measurable temporal broadening under the investigated conditions.
Additionally, the same q-plate structure is also capable of generating radially and azimuthally polarized light upon reflection of linearly polarized light.

The simple analytical description of the phase pattern allows a straightforward calculation of the nanorod arrangement for arbitrary values of OAM. At the same time, the fabrication is comparably inexpensive, and the choice of different metals may offer a new fabrication method for q-plates operating in the ultraviolet wavelength range.

\onecolumn
\section*{Author Information}
\subsection*{Author Contributions}
CW and TE wrote the original draft, prepared the visualizations, and are responsible for the data evaluation.
CW investigated the pulse duration and fabricated the nanostructure.
TE conceived the ideas, designed the experiments, performed the FDTD simulations and carried out the reflectivity measurements and beam profile investigations.
EB carried out the conversion ratio measurements.
BL contributed to writing, review, and editing, performed the e-beam lithography writing, and supported the manufacturing process.
LR, BS and MA contributed to writing, review, and editing, and were involved in project supervision.

\subsection*{Funding}
The authors gratefully acknowledge the resources on the LiCCA HPC cluster of the University of Augsburg, co-funded by the Deutsche Forschungsgemeinschaft (DFG, German Research Foundation) – Project-ID 499211671.

\subsection*{Conflict of Interest}
Authors state no conflict of interest.

\subsection*{Data Availability Statement}
The datasets generated and/or analyzed during the current study are available from the corresponding author upon reasonable request.

\subsection*{Acknowledgments}
The authors would like to thank the Nano Structure Center (NSC) at the RPTU University Kaiserslautern-Landau for sample preparation.

The authors also thank the Allianz für Hochleistungsrechnen Rheinland-Pfalz (AHRP) for providing computing resources on the High Performance Computer (HPC) \textit{Elwetritsch} at the RPTU University Kaiserslautern-Landau, which was used to generate the pattern files required for electron-beam lithography.

\printbibliography

@article{aieta2012,
  title = {Out-of-{{Plane Reflection}} and {{Refraction}} of {{Light}} by {{Anisotropic Optical Antenna Metasurfaces}} with {{Phase Discontinuities}}},
  author = {Aieta, Francesco and Genevet, Patrice and Yu, Nanfang and Kats, Mikhail A. and Gaburro, Zeno and Capasso, Federico},
  date = {2012-03-14},
  journaltitle = {Nano Letters},
  shortjournal = {Nano Lett.},
  volume = {12},
  number = {3},
  pages = {1702--1706},
  issn = {1530-6984, 1530-6992},
  doi = {10.1021/nl300204s},
  url = {https://pubs.acs.org/doi/10.1021/nl300204s},
  urldate = {2026-03-26},
}

@article{allen1992,
  title = {Orbital Angular Momentum of Light and the Transformation of {{Laguerre-Gaussian}} Laser Modes},
  author = {Allen, L. and Beijersbergen, M. W. and Spreeuw, R. J. C. and Woerdman, J. P.},
  date = {1992-06-01},
  journaltitle = {Physical Review A},
  shortjournal = {Phys. Rev. A},
  volume = {45},
  number = {11},
  pages = {8185--8189},
  publisher = {American Physical Society},
  issn = {1050-2947, 1094-1622},
  doi = {10.1103/PhysRevA.45.8185},
  url = {https://link.aps.org/doi/10.1103/PhysRevA.45.8185},
  urldate = {2026-03-24},
  langid = {english},
}

@article{arbabi2015,
  title = {Dielectric Metasurfaces for Complete Control of Phase and Polarization with Subwavelength Spatial Resolution and High Transmission},
  author = {Arbabi, Amir and Horie, Yu and Bagheri, Mahmood and Faraon, Andrei},
  date = {2015-11},
  journaltitle = {Nature Nanotechnology},
  shortjournal = {Nat. Nanotechnol.},
  volume = {10},
  number = {11},
  pages = {937--943},
  issn = {1748-3387, 1748-3395},
  doi = {10.1038/nnano.2015.186},
  url = {https://www.nature.com/articles/nnano.2015.186},
  urldate = {2026-03-27},
  langid = {english},
}

@article{bae2020,
  title = {Generation of Low-Order {{Laguerre-Gaussian}} Beams Using Hybrid-Machined Reflective Spiral Phase Plates for Intense Laser-Plasma Interactions},
  author = {Bae, Ji Yong and Jeon, Cheonha and Pae, Ki Hong and Kim, Chul Min and Kim, Hong Seung and Han, Ilkyu and Yeo, Woo-Jong and Jeong, Byeongjoon and Jeon, Minwoo and Lee, Dong-Ho and Kim, Dong Uk and Hyun, Sangwon and Hur, Hwan and Lee, Kye-Sung and Kim, Geon Hee and Chang, Ki Soo and Choi, Il Woo and Nam, Chang Hee and Kim, I Jong},
  date = {2020-12},
  journaltitle = {Results in Physics},
  shortjournal = {Results Phys.},
  volume = {19},
  pages = {103499},
  issn = {22113797},
  doi = {10.1016/j.rinp.2020.103499},
  url = {https://linkinghub.elsevier.com/retrieve/pii/S2211379720319537},
  urldate = {2026-04-28},
  langid = {english},
}

@article{beck2010,
  title = {Application of Cooled Spatial Light Modulator for High Power Nanosecond Laser Micromachining},
  author = {Beck, Rainer J and Parry, Jonathan P and MacPherson, William N and Waddie, Andrew and Weston, Nick J and Shephard, Jonathan D and Hand, Duncan P},
  date = {2010-08-02},
  journaltitle = {Optics Express},
  shortjournal = {Opt. Express},
  volume = {18},
  number = {16},
  pages = {17059--17065},
  issn = {1094-4087},
  doi = {10.1364/OE.18.017059},
  url = {https://opg.optica.org/oe/abstract.cfm?uri=oe-18-16-17059},
  urldate = {2026-04-28},
  langid = {english},
}

@article{beijersbergen1994,
  title = {Helical-Wavefront Laser Beams Produced with a Spiral Phaseplate},
  author = {Beijersbergen, M W and Coerwinkel, R P C and Kristensen, M and Woerdman, J. P.},
  date = {1994},
  journaltitle = {Optics Communications},
  shortjournal = {Opt. Commun.},
  volume = {112},
  number = {5},
  pages = {321--327},
  issn = {0030-4018},
  doi = {10.1016/0030-4018(94)90638-6},
  url = {https://www.sciencedirect.com/science/article/pii/0030401894906386},
  langid = {english},
}

@article{beth1936,
  title = {Mechanical {{Detection}} and {{Measurement}} of the {{Angular Momentum}} of {{Light}}},
  author = {Beth, Richard A.},
  date = {1936-07-15},
  journaltitle = {Physical Review},
  shortjournal = {Phys. Rev.},
  volume = {50},
  number = {2},
  pages = {115--125},
  issn = {0031-899X},
  doi = {10.1103/PhysRev.50.115},
  url = {https://link.aps.org/doi/10.1103/PhysRev.50.115},
  urldate = {2026-03-24},
  langid = {english},
}

@article{biener2002,
  title = {Formation of Helical Beams by Use of {{Pancharatnam}}--{{Berry}} Phase Optical Elements},
  author = {Biener, Gabriel and Niv, Avi and Kleiner, Vladimir and Hasman, Erez},
  date = {2002-11-01},
  journaltitle = {Optics Letters},
  shortjournal = {Opt. Lett.},
  volume = {27},
  number = {21},
  pages = {1875--1877},
  issn = {0146-9592, 1539-4794},
  doi = {10.1364/OL.27.001875},
  url = {https://opg.optica.org/abstract.cfm?URI=ol-27-21-1875},
  urldate = {2026-03-26},
  langid = {english},
}

@article{bouchon2015,
  title = {L-Shaped Metallic Antenna for Linear Polarization Conversion in Reflection},
  author = {Bouchon, Patrick and L\'evesque, Quentin and Makhsiyan, Mathilde and Pardo, Fabrice and Jaeck, Julien and Ha\"idar, Riad and Pelouard, Jean-Luc},
  date = {2015-02-27},
  journaltitle = {Photonic and Phononic Properties of Engineered Nanostructures V},
  shortjournal = {Photonic Phononic Prop. Eng. Nanostructures V},
  volume = {9371},
  pages = {93710O},
  publisher = {SPIE},
  location = {San Francisco, California, United States},
  doi = {10.1117/12.2080143},
  url = {http://proceedings.spiedigitallibrary.org/proceeding.aspx?doi=10.1117/12.2080143},
  urldate = {2026-03-27},
  langid = {english},
}

@article{bozinovic2013,
  title = {Terabit-{{Scale Orbital Angular Momentum Mode Division Multiplexing}} in {{Fibers}}},
  author = {Bozinovic, Nenad and Yue, Yang and Ren, Yongxiong and Tur, Moshe and Kristensen, Poul and Huang, Hao and Willner, Alan E. and Ramachandran, Siddharth},
  date = {2013-06-28},
  journaltitle = {Science},
  volume = {340},
  number = {6140},
  pages = {1545--1548},
  issn = {0036-8075, 1095-9203},
  doi = {10.1126/science.1237861},
  url = {https://www.science.org/doi/10.1126/science.1237861},
  urldate = {2025-09-19},
  langid = {english},
}

@article{campbell2012,
  title = {Generation of High-Order Optical Vortices Using Directly Machined Spiral Phase Mirrors},
  author = {Campbell, Geoff and Hage, Boris and Buchler, Ben and Lam, Ping Koy},
  date = {2012-03-01},
  journaltitle = {Applied Optics},
  shortjournal = {Appl. Opt.},
  volume = {51},
  number = {7},
  pages = {873--876},
  issn = {1559-128X, 2155-3165},
  doi = {10.1364/AO.51.000873},
  url = {https://opg.optica.org/abstract.cfm?URI=ao-51-7-873},
  urldate = {2026-04-28},
  langid = {english},
}

@article{chen2016,
  title = {A Review of Metasurfaces: Physics and Applications},
  shorttitle = {A Review of Metasurfaces},
  author = {Chen, Hou-Tong and Taylor, Antoinette J and Yu, Nanfang},
  date = {2016-07-01},
  journaltitle = {Reports on Progress in Physics},
  shortjournal = {Rep. Prog. Phys.},
  volume = {79},
  number = {7},
  pages = {076401},
  issn = {0034-4885, 1361-6633},
  doi = {10.1088/0034-4885/79/7/076401},
  url = {https://iopscience.iop.org/article/10.1088/0034-4885/79/7/076401},
  urldate = {2026-03-27},
  langid = {english},
}

@article{chen2016a,
  title = {Artificial Perfect Electric Conductor-Perfect Magnetic Conductor Anisotropic Metasurface for Generating Orbital Angular Momentum of Microwave with Nearly Perfect Conversion Efficiency},
  author = {Chen, Menglin L. N. and Jiang, Li Jun and Sha, Wei E. I.},
  date = {2016-02-14},
  journaltitle = {Journal of Applied Physics},
  shortjournal = {J. Appl. Phys.},
  volume = {119},
  number = {6},
  pages = {064506},
  issn = {0021-8979, 1089-7550},
  doi = {10.1063/1.4941696},
  url = {https://pubs.aip.org/jap/article/119/6/064506/142625/Artificial-perfect-electric-conductor-perfect},
  urldate = {2026-08-28},
  langid = {english},
}

@article{chen2018,
  title = {Orbital {{Angular Momentum Generation}} and {{Detection}} by {{Geometric-Phase Based Metasurfaces}}},
  author = {Chen, Menglin and Jiang, Li and Sha, Wei},
  date = {2018-03-02},
  journaltitle = {Applied Sciences},
  shortjournal = {Appl. Sci.},
  volume = {8},
  number = {3},
  pages = {362},
  issn = {2076-3417},
  doi = {10.3390/app8030362},
  url = {https://www.mdpi.com/2076-3417/8/3/362},
  urldate = {2026-03-27},
  langid = {english},
}

@article{denisenko2009,
  title = {Determination of Topological Charges of Polychromatic Optical Vortices},
  author = {Denisenko, Vladimir and Shvedov, Vladlen and Desyatnikov, Anton S. and Neshev, Dragomir N. and Krolikowski, Wieslaw and Volyar, Alexander and Soskin, Marat and Kivshar, Yuri S.},
  date = {2009-12-21},
  journaltitle = {Optics Express},
  shortjournal = {Opt. Express},
  volume = {17},
  number = {26},
  pages = {23374--23379},
  issn = {1094-4087},
  doi = {10.1364/OE.17.023374},
  url = {https://opg.optica.org/oe/abstract.cfm?uri=oe-17-26-23374},
  urldate = {2026-03-05},
  langid = {english},
}

@article{ding2015,
  title = {Broadband {{High-Efficiency Half-Wave Plate}}: {{A Supercell-Based Plasmonic Metasurface Approach}}},
  shorttitle = {Broadband {{High-Efficiency Half-Wave Plate}}},
  author = {Ding, Fei and Wang, Zhuoxian and He, Sailing and Shalaev, Vladimir M. and Kildishev, Alexander V.},
  date = {2015-04-28},
  journaltitle = {ACS Nano},
  volume = {9},
  number = {4},
  pages = {4111--4119},
  issn = {1936-0851, 1936-086X},
  doi = {10.1021/acsnano.5b00218},
  url = {https://pubs.acs.org/doi/10.1021/acsnano.5b00218},
  urldate = {2023-08-03},
  langid = {english},
}

@article{ding2020,
  title = {Focused Vortex-beam Generation Using Gap-surface Plasmon Metasurfaces},
  author = {Ding, Fei and Chen, Yiting and Bozhevolnyi, Sergey I.},
  date = {2020-02},
  journaltitle = {Nanophotonics},
  volume = {9},
  number = {2},
  pages = {371--378},
  issn = {2192-8614, 2192-8614},
  doi = {10.1515/nanoph-2019-0235},
  url = {https://onlinelibrary.wiley.com/doi/10.1515/nanoph-2019-0235},
  urldate = {2026-08-28},
  langid = {english},
}

@article{ebel2023,
  title = {Optical Reflective Metasurfaces Based on Mirror-Coupled Slot Antennas},
  author = {Ebel, Sven and Deng, Yadong and Hentschel, Mario and Meng, Chao and Sande, S\"oren Im and Giessen, Harald and Ding, Fei and Bozhevolnyi, Sergey I.},
  date = {2023-01-02},
  journaltitle = {Advanced Photonics Nexus},
  shortjournal = {Adv. Photonics Nexus},
  volume = {2},
  number = {1},
  pages = {016005},
  issn = {2791-1519},
  doi = {10.1117/1.APN.2.1.016005},
  url = {https://www.spiedigitallibrary.org/journals/advanced-photonics-nexus/volume-2/issue-01/016005/Optical-reflective-metasurfaces-based-on-mirror-coupled-slot-antennas/10.1117/1.APN.2.1.016005.full},
  urldate = {2026-04-29},
  langid = {english},
}

@article{efimov1995,
  title = {Programmable Shaping of Ultrabroad-Bandwidth Pulses from a {{Ti}}:Sapphire Laser},
  shorttitle = {Programmable Shaping of Ultrabroad-Bandwidth Pulses from a {{Ti}}},
  author = {Efimov, A. and Schaffer, C. and Reitze, D. H.},
  date = {1995-10-01},
  journaltitle = {Journal of the Optical Society of America B},
  shortjournal = {J. Opt. Soc. Am. B},
  volume = {12},
  number = {10},
  pages = {1968--1980},
  issn = {0740-3224, 1520-8540},
  doi = {10.1364/JOSAB.12.001968},
  url = {https://opg.optica.org/abstract.cfm?URI=josab-12-10-1968},
  urldate = {2026-03-25},
  langid = {english},
}

@article{elliott2004,
  title = {Wavelength Dependent Birefringence of Surface Plasmon Polaritonic Crystals},
  author = {Elliott, Jill and Smolyaninov, Igor I. and Zheludev, Nikolay I. and Zayats, Anatoly V.},
  date = {2004-12-06},
  journaltitle = {Physical Review B},
  shortjournal = {Phys. Rev. B},
  volume = {70},
  number = {23},
  pages = {233403},
  issn = {1098-0121, 1550-235X},
  doi = {10.1103/PhysRevB.70.233403},
  url = {https://link.aps.org/doi/10.1103/PhysRevB.70.233403},
  urldate = {2026-03-26},
  langid = {english},
}

@article{erhard2017,
  title = {Twisted Photons: New Quantum Perspectives in High Dimensions},
  shorttitle = {Twisted Photons},
  author = {Erhard, Manuel and Fickler, Robert and Krenn, Mario and Zeilinger, Anton},
  date = {2017-10-17},
  journaltitle = {Light: Science \& Applications},
  shortjournal = {Light Sci. Appl.},
  volume = {7},
  number = {3},
  pages = {17146--17146},
  issn = {2047-7538},
  doi = {10.1038/lsa.2017.146},
  url = {https://www.nature.com/articles/lsa2017146},
  urldate = {2026-04-22},
  langid = {english},
}

@article{forbes2016,
  title = {Creation and Detection of Optical Modes with Spatial Light Modulators},
  author = {Forbes, Andrew and Dudley, Angela and McLaren, Melanie},
  date = {2016-06-30},
  journaltitle = {Advances in Optics and Photonics},
  shortjournal = {Adv. Opt. Photonics},
  volume = {8},
  number = {2},
  pages = {200},
  issn = {1943-8206},
  doi = {10.1364/AOP.8.000200},
  url = {https://opg.optica.org/abstract.cfm?URI=aop-8-2-200},
  urldate = {2026-03-25},
  langid = {english},
}

@article{georgi2017,
  title = {Rotational {{Doppler}} Shift Induced by Spin-Orbit Coupling of Light at Spinning Metasurfaces},
  author = {Georgi, Philip and Schlickriede, Christian and Li, Guixin and Zhang, Shuang and Zentgraf, Thomas},
  date = {2017-08-20},
  journaltitle = {Optica},
  volume = {4},
  number = {8},
  pages = {1000--1005},
  issn = {2334-2536},
  doi = {10.1364/OPTICA.4.001000},
  url = {https://opg.optica.org/abstract.cfm?URI=optica-4-8-1000},
  urldate = {2026-08-28},
  langid = {english},
}

@article{grady2013,
  title = {Terahertz {{Metamaterials}} for {{Linear Polarization Conversion}} and {{Anomalous Refraction}}},
  author = {Grady, Nathaniel K. and Heyes, Jane E. and Chowdhury, Dibakar Roy and Zeng, Yong and Reiten, Matthew T. and Azad, Abul K. and Taylor, Antoinette J. and Dalvit, Diego A. R. and Chen, Hou-Tong},
  date = {2013-06-14},
  journaltitle = {Science},
  volume = {340},
  number = {6138},
  pages = {1304--1307},
  issn = {0036-8075, 1095-9203},
  doi = {10.1126/science.1235399},
  url = {https://www.science.org/doi/10.1126/science.1235399},
  urldate = {2024-01-15},
  langid = {english},
}

@article{he1995,
  title = {Optical {{Particle Trapping}} with {{Higher-order Doughnut Beams Produced Using High Efficiency Computer Generated Holograms}}},
  author = {He, H. and Heckenberg, N.R. and Rubinsztein-Dunlop, H.},
  date = {1995-01},
  journaltitle = {Journal of Modern Optics},
  shortjournal = {J. Mod. Opt.},
  volume = {42},
  number = {1},
  pages = {217--223},
  issn = {0950-0340, 1362-3044},
  doi = {10.1080/09500349514550171},
  url = {http://www.tandfonline.com/doi/abs/10.1080/09500349514550171},
  urldate = {2026-04-22},
  langid = {english},
}

@article{heiden2019,
  title = {Gap-{{Surface Plasmon Metasurfaces}} for {{Broadband Circular}}-to-{{Linear Polarization Conversion}} and {{Vector Vortex Beam Generation}}},
  author = {Heiden, Jacob T. and Ding, Fei and Linnet, Jes and Yang, Yuanqing and Beermann, Jonas and Bozhevolnyi, Sergey I.},
  date = {2019-05},
  journaltitle = {Advanced Optical Materials},
  shortjournal = {Adv. Opt. Mater.},
  volume = {7},
  number = {9},
  pages = {1801414},
  issn = {2195-1071, 2195-1071},
  doi = {10.1002/adom.201801414},
  url = {https://advanced.onlinelibrary.wiley.com/doi/10.1002/adom.201801414},
  urldate = {2026-08-28},
  langid = {english},
}

@article{huang2007,
  title = {Plasmonic Optical Properties of a Single Gold Nano-Rod},
  author = {Huang, Hung Ji and Yu, Chin-ping and Chang, Hung Chun and Chiu, Kuo Pin and Ming Chen, Hao and Liu, Ru Shi and Tsai, Din Ping},
  date = {2007},
  journaltitle = {Optics Express},
  shortjournal = {Opt. Express},
  volume = {15},
  number = {12},
  pages = {7132--7139},
  issn = {1094-4087},
  doi = {10.1364/OE.15.007132},
  url = {https://opg.optica.org/oe/abstract.cfm?uri=oe-15-12-7132},
  urldate = {2026-03-27},
  langid = {english},
}

@article{huang2012,
  title = {Dispersionless {{Phase Discontinuities}} for {{Controlling Light Propagation}}},
  author = {Huang, Lingling and Chen, Xianzhong and M\"uhlenbernd, Holger and Li, Guixin and Bai, Benfeng and Tan, Qiaofeng and Jin, Guofan and Zentgraf, Thomas and Zhang, Shuang},
  date = {2012-11-14},
  journaltitle = {Nano Letters},
  shortjournal = {Nano Lett.},
  volume = {12},
  number = {11},
  pages = {5750--5755},
  issn = {1530-6984, 1530-6992},
  doi = {10.1021/nl303031j},
  url = {https://pubs.acs.org/doi/10.1021/nl303031j},
  urldate = {2026-03-27},
  langid = {english},
}

@article{jiang2014,
  title = {Broadband and {{Wide Field-of-view Plasmonic Metasurface-enabled Waveplates}}},
  author = {Jiang, Zhi Hao and Lin, Lan and Ma, Ding and Yun, Seokho and Werner, Douglas H. and Liu, Zhiwen and Mayer, Theresa S.},
  date = {2014-12-19},
  journaltitle = {Scientific Reports},
  shortjournal = {Sci. Rep.},
  volume = {4},
  number = {1},
  pages = {7511},
  issn = {2045-2322},
  doi = {10.1038/srep07511},
  url = {https://www.nature.com/articles/srep07511},
  urldate = {2023-08-03},
  langid = {english},
}

@article{jullien2016,
  title = {Continuously Tunable Femtosecond Delay-Line Based on Liquid Crystal Cells},
  author = {Jullien, Aur\'elie and Bortolozzo, Umberto and Grabielle, St\'ephanie and Huignard, Jean-Pierre and Forget, Nicolas and Residori, Stefania},
  date = {2016-06-27},
  journaltitle = {Optics Express},
  shortjournal = {Opt. Express},
  volume = {24},
  number = {13},
  pages = {14483--14493},
  issn = {1094-4087},
  doi = {10.1364/OE.24.014483},
  url = {https://opg.optica.org/abstract.cfm?URI=oe-24-13-14483},
  urldate = {2026-03-26},
  langid = {english},
}

@article{karimi2014,
  title = {Generating Optical Orbital Angular Momentum at Visible Wavelengths Using a Plasmonic Metasurface},
  author = {Karimi, Ebrahim and Schulz, Sebastian A and De Leon, Israel and Qassim, Hammam and Upham, Jeremy and Boyd, Robert W},
  date = {2014-05-09},
  journaltitle = {Light: Science \& Applications},
  shortjournal = {Light Sci. Appl.},
  volume = {3},
  number = {5},
  pages = {e167-e167},
  issn = {2047-7538},
  doi = {10.1038/lsa.2014.48},
  url = {https://www.nature.com/articles/lsa201448},
  urldate = {2026-03-27},
  langid = {english},
}

@article{kats2012,
  title = {Giant Birefringence in Optical Antenna Arrays with Widely Tailorable Optical Anisotropy},
  author = {Kats, Mikhail A. and Genevet, Patrice and Aoust, Guillaume and Yu, Nanfang and Blanchard, Romain and Aieta, Francesco and Gaburro, Zeno and Capasso, Federico},
  date = {2012-07-31},
  journaltitle = {Proceedings of the National Academy of Sciences},
  shortjournal = {Proc. Natl. Acad. Sci.},
  volume = {109},
  number = {31},
  pages = {12364--12368},
  issn = {0027-8424, 1091-6490},
  doi = {10.1073/pnas.1210686109},
  url = {https://pnas.org/doi/full/10.1073/pnas.1210686109},
  urldate = {2026-03-26},
  langid = {english},
}

@article{konforti1988,
  title = {Phase-Only Modulation with Twisted Nematic Liquid-Crystal Spatial Light Modulators},
  author = {Konforti, N. and Marom, E. and Wu, S.-T.},
  date = {1988-03-01},
  journaltitle = {Optics Letters},
  shortjournal = {Opt. Lett.},
  volume = {13},
  number = {3},
  pages = {251--253},
  issn = {0146-9592, 1539-4794},
  doi = {10.1364/OL.13.000251},
  url = {https://opg.optica.org/ol/abstract.cfm?uri=ol-13-3-251},
  urldate = {2026-03-25},
  langid = {english},
}

@article{link1999,
  title = {Simulation of the {{Optical Absorption Spectra}} of {{Gold Nanorods}} as a {{Function}} of {{Their Aspect Ratio}} and the {{Effect}} of the {{Medium Dielectric Constant}}},
  author = {Link, S. and Mohamed, M. B. and El-Sayed, M. A.},
  date = {1999-04-01},
  journaltitle = {The Journal of Physical Chemistry B},
  shortjournal = {J. Phys. Chem. B},
  volume = {103},
  number = {16},
  pages = {3073--3077},
  publisher = {American Chemical Society},
  issn = {1520-6106, 1520-5207},
  doi = {10.1021/jp990183f},
  url = {https://pubs.acs.org/doi/10.1021/jp990183f},
  urldate = {2026-03-27},
  langid = {english},
}

@article{liu2017,
  title = {Single-{{Layer Plasmonic Metasurface Half-Wave Plates}} with {{Wavelength-Independent Polarization Conversion Angle}}},
  author = {Liu, Zhaocheng and Li, Zhancheng and Liu, Zhe and Cheng, Hua and Liu, Wenwei and Tang, Chengchun and Gu, Changzhi and Li, Junjie and Chen, Hou-Tong and Chen, Shuqi and Tian, Jianguo},
  date = {2017-08-16},
  journaltitle = {ACS Photonics},
  volume = {4},
  number = {8},
  pages = {2061--2069},
  issn = {2330-4022, 2330-4022},
  doi = {10.1021/acsphotonics.7b00491},
  url = {https://pubs.acs.org/doi/10.1021/acsphotonics.7b00491},
  urldate = {2023-08-03},
  langid = {english},
}

@article{longman2020,
  title = {Off-Axis Spiral Phase Mirrors for Generating High-Intensity Optical Vortices},
  author = {Longman, Andrew and Salgado, Carlos and Zeraouli, Ghassan and Api\~naniz, Jon I. and Antonio P\'erez-Hern\'andez, Jose and Eltahlawy, M. Khairy and Volpe, Luca and Fedosejevs, Robert},
  date = {2020-04-15},
  journaltitle = {Optics Letters},
  shortjournal = {Opt. Lett.},
  volume = {45},
  number = {8},
  pages = {2187--2190},
  issn = {0146-9592, 1539-4794},
  doi = {10.1364/OL.387363},
  url = {https://opg.optica.org/abstract.cfm?URI=ol-45-8-2187},
  urldate = {2026-04-28},
  langid = {english},
}

@article{marrucci2006,
  title = {Optical {{Spin-to-Orbital Angular Momentum Conversion}} in {{Inhomogeneous Anisotropic Media}}},
  author = {Marrucci, L. and Manzo, C. and Paparo, D.},
  date = {2006-04-28},
  journaltitle = {Physical Review Letters},
  shortjournal = {Phys. Rev. Lett.},
  volume = {96},
  number = {16},
  pages = {163905},
  issn = {0031-9007, 1079-7114},
  doi = {10.1103/PhysRevLett.96.163905},
  url = {https://link.aps.org/doi/10.1103/PhysRevLett.96.163905},
  urldate = {2023-10-25},
  langid = {english},
}

@article{marrucci2012,
  title = {Spin-to-{{Orbital Optical Angular Momentum Conversion}} in {{Liquid Crystal}} ``q-{{Plates}}'': {{Classical}} and {{Quantum Applications}}},
  shorttitle = {Spin-to-{{Orbital Optical Angular Momentum Conversion}} in {{Liquid Crystal}} ``q-{{Plates}}''},
  author = {Marrucci, L. and Karimi, E. and Slussarenko, S. and Piccirillo, B. and Santamato, E. and Nagali, E. and Sciarrino, F.},
  date = {2012-08-17},
  journaltitle = {Molecular Crystals and Liquid Crystals},
  shortjournal = {Mol. Cryst. Liq. Cryst.},
  volume = {561},
  number = {1},
  pages = {48--56},
  issn = {1542-1406, 1563-5287},
  doi = {10.1080/15421406.2012.686710},
  url = {http://www.tandfonline.com/doi/abs/10.1080/15421406.2012.686710},
  urldate = {2026-03-27},
  langid = {english},
}

@article{oemrawsingh2004,
  title = {Production and Characterization of Spiral Phase Plates for Optical Wavelengths},
  author = {Oemrawsingh, S. S. R. and Van Houwelingen, J. A. W. and Eliel, E. R. and Woerdman, J. P. and Verstegen, E. J. K. and Kloosterboer, J. G. and 'T Hooft, G. W.},
  date = {2004-01-20},
  journaltitle = {Applied Optics},
  shortjournal = {Appl. Opt.},
  volume = {43},
  number = {3},
  pages = {688--694},
  issn = {0003-6935, 1539-4522},
  doi = {10.1364/AO.43.000688},
  url = {https://opg.optica.org/abstract.cfm?URI=ao-43-3-688},
  urldate = {2026-03-25},
  langid = {english},
}

@article{olson2015,
  title = {Optical Characterization of Single Plasmonic Nanoparticles},
  author = {Olson, Jana and Dominguez-Medina, Sergio and Hoggard, Anneli and Wang, Lin-Yung and Chang, Wei-Shun and Link, Stephan},
  date = {2015-01-07},
  journaltitle = {Chemical Society Reviews},
  shortjournal = {Chem. Soc. Rev.},
  volume = {44},
  number = {1},
  pages = {40--57},
  issn = {0306-0012, 1460-4744},
  doi = {10.1039/C4CS00131A},
  url = {https://xlink.rsc.org/?DOI=C4CS00131A},
  urldate = {2026-03-27},
  langid = {english},
}

@article{padgett2004,
  title = {Light's {{Orbital Angular Momentum}}},
  author = {Padgett, Miles and Courtial, Johannes and Allen, Les},
  date = {2004-05-01},
  journaltitle = {Physics Today},
  shortjournal = {Phys. Today},
  volume = {57},
  number = {5},
  pages = {35--40},
  issn = {0031-9228, 1945-0699},
  doi = {10.1063/1.1768672},
  url = {https://physicstoday.aip.org/features/lights-orbital-angular-momentum},
  urldate = {2025-10-28},
  langid = {english},
}

@article{padgett2011,
  title = {Tweezers with a Twist},
  author = {Padgett, Miles and Bowman, Richard},
  date = {2011-06-01},
  journaltitle = {Nature Photonics},
  shortjournal = {Nat. Photonics},
  volume = {5},
  number = {6},
  pages = {343--348},
  issn = {1749-4885, 1749-4893},
  doi = {10.1038/nphoton.2011.81},
  url = {https://www.nature.com/articles/nphoton.2011.81},
  urldate = {2026-04-22},
  langid = {english},
}

@article{pors2013,
  title = {Efficient and Broadband Quarter-Wave Plates by Gap-Plasmon Resonators},
  author = {Pors, Anders and Bozhevolnyi, Sergey I.},
  date = {2013-02-11},
  journaltitle = {Optics Express},
  shortjournal = {Opt. Express},
  volume = {21},
  number = {3},
  pages = {2942--2952},
  issn = {1094-4087},
  doi = {10.1364/OE.21.002942},
  url = {https://opg.optica.org/oe/abstract.cfm?uri=oe-21-3-2942},
  urldate = {2026-03-27},
  langid = {english},
}

@article{pors2013a,
  title = {Plasmonic Metasurfaces for Efficient Phase Control in Reflection},
  author = {Pors, Anders and Bozhevolnyi, Sergey I.},
  date = {2013-11-04},
  journaltitle = {Optics Express},
  shortjournal = {Opt. Express},
  volume = {21},
  number = {22},
  pages = {27438--27451},
  issn = {1094-4087},
  doi = {10.1364/OE.21.027438},
  url = {https://opg.optica.org/oe/abstract.cfm?uri=oe-21-22-27438},
  urldate = {2026-03-26},
  langid = {english},
}

@article{pors2013b,
  title = {Broadband Plasmonic Half-Wave Plates in Reflection},
  author = {Pors, Anders and Nielsen, Michael G. and Bozhevolnyi, Sergey I.},
  date = {2013-02-15},
  journaltitle = {Optics Letters},
  shortjournal = {Opt. Lett.},
  volume = {38},
  number = {4},
  pages = {513--515},
  issn = {0146-9592, 1539-4794},
  doi = {10.1364/OL.38.000513},
  url = {https://opg.optica.org/abstract.cfm?URI=ol-38-4-513},
  urldate = {2026-03-27},
  langid = {english},
}

@article{poynting1909,
  title = {The {{Wave Motion}} of a {{Revolving Shaft}}, and a {{Suggestion}} as to the {{Angular Momentum}} in a {{Beam}} of {{Circularly Polarised Light}}},
  author = {Poynting, John Henry},
  date = {1909-07-31},
  journaltitle = {Proceedings of the Royal Society of London. Series A, Containing Papers of a Mathematical and Physical Character},
  shortjournal = {Proc. R. Soc. Lond. Ser. Contain. Pap. Math. Phys. Character},
  volume = {82},
  number = {557},
  pages = {560--567},
  issn = {0950-1207},
  doi = {10.1098/rspa.1909.0060},
  url = {https://doi.org/10.1098/rspa.1909.0060},
  langid = {english},
}

@article{rafayelyan2016,
  title = {Bragg-{{Berry}} Mirrors: Reflective Broadband q-Plates},
  shorttitle = {Bragg-{{Berry}} Mirrors},
  author = {Rafayelyan, Mushegh and Brasselet, Etienne},
  date = {2016-09-01},
  journaltitle = {Optics Letters},
  shortjournal = {Opt. Lett.},
  volume = {41},
  number = {17},
  pages = {3972--3975},
  issn = {0146-9592, 1539-4794},
  doi = {10.1364/OL.41.003972},
  url = {https://opg.optica.org/abstract.cfm?URI=ol-41-17-3972},
  urldate = {2026-03-26},
  langid = {english},
}

@article{rubano2019,
  title = {Q-Plate Technology: A Progress Review [{{Invited}}]},
  shorttitle = {Q-Plate Technology},
  author = {Rubano, Andrea and Cardano, Filippo and Piccirillo, Bruno and Marrucci, Lorenzo},
  date = {2019-05-01},
  journaltitle = {Journal of the Optical Society of America B},
  shortjournal = {J. Opt. Soc. Am. B},
  volume = {36},
  number = {5},
  pages = {D70-D87},
  issn = {0740-3224, 1520-8540},
  doi = {10.1364/JOSAB.36.000D70},
  url = {https://opg.optica.org/abstract.cfm?URI=josab-36-5-D70},
  urldate = {2026-03-26},
  langid = {english},
}

@article{sanchez-lopez2018,
  title = {Spectral Performance of a Zero-Order Liquid-Crystal Polymer Commercial q-Plate for the Generation of Vector Beams at Different Wavelengths},
  author = {S\'anchez-L\'opez, Mar\'ia M. and Abella, Isaiah and Puerto-Garc\'ia, Daniel and Davis, Jeffrey A. and Moreno, Ignacio},
  date = {2018-10},
  journaltitle = {Optics \& Laser Technology},
  shortjournal = {Opt. Laser Technol.},
  volume = {106},
  pages = {168--176},
  issn = {0030-3992},
  doi = {10.1016/j.optlastec.2018.04.008},
  url = {https://linkinghub.elsevier.com/retrieve/pii/S0030399217316250},
  urldate = {2026-03-26},
  langid = {english},
}

@article{schmiegelow2016,
  title = {Transfer of Optical Orbital Angular Momentum to a Bound Electron},
  author = {Schmiegelow, Christian T. and Schulz, Jonas and Kaufmann, Henning and Ruster, Thomas and Poschinger, Ulrich G. and Schmidt-Kaler, Ferdinand},
  date = {2016-10-03},
  journaltitle = {Nature Communications},
  shortjournal = {Nat. Commun.},
  volume = {7},
  number = {1},
  pages = {12998},
  issn = {2041-1723},
  doi = {10.1038/ncomms12998},
  url = {https://www.nature.com/articles/ncomms12998},
  urldate = {2026-04-22},
  langid = {english},
}

@article{shen2019,
  title = {Optical Vortices 30 Years on: {{OAM}} Manipulation from Topological Charge to Multiple Singularities},
  shorttitle = {Optical Vortices 30 Years On},
  author = {Shen, Yijie and Wang, Xuejiao and Xie, Zhenwei and Min, Changjun and Fu, Xing and Liu, Qiang and Gong, Mali and Yuan, Xiaocong},
  date = {2019-10-02},
  journaltitle = {Light: Science \& Applications},
  shortjournal = {Light Sci. Appl.},
  volume = {8},
  number = {1},
  pages = {90},
  issn = {2047-7538},
  doi = {10.1038/s41377-019-0194-2},
  url = {https://www.nature.com/articles/s41377-019-0194-2},
  urldate = {2026-03-25},
  langid = {english},
}

@article{sueda2004,
  title = {Laguerre-{{Gaussian}} Beam Generated with a Multilevel Spiral Phase Plate for High Intensity Laser Pulses},
  author = {Sueda, K and Miyaji, G and Miyanaga, N and Nakatsuka, M},
  date = {2004-07},
  volume = {12},
  number = {15},
  pages = {3548--3553},
  publisher = {Optics Express},
  doi = {10.1364/OPEX.12.003548},
  url = {https://opg.optica.org/oe/abstract.cfm?URI=oe-12-15-3548},
  langid = {english},
}

@article{sung2008,
  title = {Nanoparticle {{Spectroscopy}}: {{Birefringence}} in {{Two-Dimensional Arrays}} of {{L-Shaped Silver Nanoparticles}}},
  shorttitle = {Nanoparticle {{Spectroscopy}}},
  author = {Sung, Jiha and Sukharev, Maxim and Hicks, Erin M. and Van Duyne, Richard P. and Seideman, Tamar and Spears, Kenneth G.},
  date = {2008-03-01},
  journaltitle = {The Journal of Physical Chemistry C},
  shortjournal = {J. Phys. Chem. C},
  volume = {112},
  number = {9},
  pages = {3252--3260},
  issn = {1932-7447, 1932-7455},
  doi = {10.1021/jp077389y},
  url = {https://pubs.acs.org/doi/10.1021/jp077389y},
  urldate = {2026-03-26},
  langid = {english},
}

@article{tang2020,
  title = {High-Efficiency Focused Optical Vortex Generation with Geometric Gap-Surface Plasmon Metalenses},
  author = {Tang, Shiwei and Ding, Fei},
  date = {2020-07-06},
  journaltitle = {Applied Physics Letters},
  shortjournal = {Appl. Phys. Lett.},
  volume = {117},
  number = {1},
  pages = {011103},
  issn = {0003-6951, 1077-3118},
  doi = {10.1063/5.0014822},
  url = {https://pubs.aip.org/apl/article/117/1/011103/39036/High-efficiency-focused-optical-vortex-generation},
  urldate = {2026-08-28},
  langid = {english},
}

@article{walmsness2019,
  title = {Optical Response of Rectangular Array of Elliptical Plasmonic Particles on Glass Revealed by {{Mueller}} Matrix Ellipsometry and {{Finite Element}} Modelling},
  author = {Walmsness, Per Magnus and Brakstad, Thomas and Svendsen, Brage B and Banon, Jean-Philippe and Walmsley, John C and Kildemo, Morten},
  date = {2019-07},
  journaltitle = {Journal of the Optical Society of America B},
  shortjournal = {J. Opt. Soc. Am. B},
  volume = {36},
  number = {7},
  pages = {E78-E87},
  publisher = {Optica Publishing Group},
  doi = {10.1364/JOSAB.36.000E78},
  url = {https://opg.optica.org/josab/abstract.cfm?URI=josab-36-7-E78},
  langid = {english},
}

@article{wang2012,
  title = {Polarization Conversion with Elliptical Patch Nanoantennas},
  author = {Wang, Feng and Chakrabarty, Ayan and Minkowski, Fred and Sun, Kai and Wei, Qi-Huo},
  date = {2012-07-09},
  journaltitle = {Applied Physics Letters},
  shortjournal = {Appl. Phys. Lett.},
  volume = {101},
  number = {2},
  pages = {023101},
  issn = {0003-6951, 1077-3118},
  doi = {10.1063/1.4731792},
  url = {https://pubs.aip.org/apl/article/101/2/023101/128097/Polarization-conversion-with-elliptical-patch},
  urldate = {2026-03-27},
  langid = {english},
}

@article{wang2012a,
  title = {Terabit Free-Space Data Transmission Employing Orbital Angular Momentum Multiplexing},
  author = {Wang, Jian and Yang, Jeng-Yuan and Fazal, Irfan M. and Ahmed, Nisar and Yan, Yan and Huang, Hao and Ren, Yongxiong and Yue, Yang and Dolinar, Samuel and Tur, Moshe and Willner, Alan E.},
  date = {2012-07},
  journaltitle = {Nature Photonics},
  shortjournal = {Nat. Photonics},
  volume = {6},
  number = {7},
  pages = {488--496},
  issn = {1749-4885, 1749-4893},
  doi = {10.1038/nphoton.2012.138},
  url = {https://www.nature.com/articles/nphoton.2012.138},
  urldate = {2025-09-19},
  langid = {english},
}

@article{wang2015,
  title = {L-Shaped Metasurface for Both the Linear and Circular Polarization Conversions},
  author = {Wang, Wei and Guo, Zhongyi and Li, Rongzhen and Zhang, Jingran and Zhang, Anjun and Li, Yan and Liu, Yi and Wang, Xinshun and Qu, Shiliang},
  date = {2015-06-01},
  journaltitle = {Journal of Optics},
  shortjournal = {J. Opt.},
  volume = {17},
  number = {6},
  pages = {065103},
  issn = {2040-8978, 2040-8986},
  doi = {10.1088/2040-8978/17/6/065103},
  url = {https://iopscience.iop.org/article/10.1088/2040-8978/17/6/065103},
  urldate = {2026-03-27},
  langid = {english},
}

@article{wang2015a,
  title = {Quantum Teleportation of Multiple Degrees of Freedom of a Single Photon},
  author = {Wang, Xi-Lin and Cai, Xin-Dong and Su, Zu-En and Chen, Ming-Cheng and Wu, Dian and Li, Li and Liu, Nai-Le and Lu, Chao-Yang and Pan, Jian-Wei},
  date = {2015-02-26},
  journaltitle = {Nature},
  volume = {518},
  number = {7540},
  pages = {516--519},
  issn = {0028-0836, 1476-4687},
  doi = {10.1038/nature14246},
  url = {https://www.nature.com/articles/nature14246},
  urldate = {2026-04-22},
  langid = {english},
}

@article{wang2016a,
  title = {Advances in Communications Using Optical Vortices},
  author = {Wang, Jian},
  date = {2016-10-01},
  journaltitle = {Photonics Research},
  shortjournal = {Photonics Res.},
  volume = {4},
  number = {5},
  pages = {B14},
  issn = {2327-9125},
  doi = {10.1364/PRJ.4.000B14},
  url = {https://opg.optica.org/abstract.cfm?URI=prj-4-5-B14},
  urldate = {2026-04-22},
  langid = {english},
}

@article{wefers1995,
  title = {Analysis of Programmable Ultrashort Waveform Generation Using Liquid-Crystal Spatial Light Modulators},
  author = {Wefers, Marc M. and Nelson, Keith A.},
  date = {1995-07-01},
  journaltitle = {Journal of the Optical Society of America B},
  shortjournal = {J. Opt. Soc. Am. B},
  volume = {12},
  number = {7},
  pages = {1343--1362},
  issn = {0740-3224, 1520-8540},
  doi = {10.1364/JOSAB.12.001343},
  url = {https://opg.optica.org/abstract.cfm?URI=josab-12-7-1343},
  urldate = {2026-03-25},
  langid = {english},
}

@article{weiss2026,
  title = {Toward {{Plasmonic Neuronal Architectures}} at the {{Nanometer Scale}}},
  author = {Wei\ss, Christopher G. O. and Eul, Tobias and Kruel, Emily and Pfeiffer, Mario F. and L\"agel, Bert and Stadtm\"uller, Benjamin and Aeschlimann, Martin},
  date = {2026-04-13},
  journaltitle = {Nanophotonics},
  volume = {15},
  number = {7},
  pages = {e70066},
  issn = {2192-8614},
  doi = {10.1002/nap2.70066},
  url = {https://onlinelibrary.wiley.com/doi/10.1002/nap2.70066},
  urldate = {2026-04-10},
  langid = {english},
}

@article{yamane2012,
  title = {Ultrashort Optical-Vortex Pulse Generation in Few-Cycle Regime},
  author = {Yamane, Keisaku and Toda, Yasunori and Morita, Ryuji},
  date = {2012-08-13},
  journaltitle = {Optics Express},
  shortjournal = {Opt. Express},
  volume = {20},
  number = {17},
  pages = {18986--18993},
  issn = {1094-4087},
  doi = {10.1364/OE.20.018986},
  url = {https://opg.optica.org/oe/abstract.cfm?uri=oe-20-17-18986},
  urldate = {2026-03-25},
  langid = {english},
}

@article{yang2019,
  title = {All-Dielectric Reflective Metasurface for Orbital Angular Momentum Beam Generation},
  author = {Yang, Jiaqi and Zhou, Heng and Lan, Tian},
  date = {2019-09-01},
  journaltitle = {Optical Materials Express},
  shortjournal = {Opt. Mater. Express},
  volume = {9},
  number = {9},
  pages = {3594--3603},
  issn = {2159-3930},
  doi = {10.1364/OME.9.003594},
  url = {https://opg.optica.org/abstract.cfm?URI=ome-9-9-3594},
  urldate = {2026-08-28},
  langid = {english},
}

@article{yu2011,
  title = {Light {{Propagation}} with {{Phase Discontinuities}}: {{Generalized Laws}} of {{Reflection}} and {{Refraction}}},
  shorttitle = {Light {{Propagation}} with {{Phase Discontinuities}}},
  author = {Yu, Nanfang and Genevet, Patrice and Kats, Mikhail A. and Aieta, Francesco and Tetienne, Jean-Philippe and Capasso, Federico and Gaburro, Zeno},
  date = {2011-10-21},
  journaltitle = {Science},
  volume = {334},
  number = {6054},
  pages = {333--337},
  issn = {0036-8075, 1095-9203},
  doi = {10.1126/science.1210713},
  url = {https://www.science.org/doi/10.1126/science.1210713},
  urldate = {2026-03-26},
  langid = {english},
}

@article{yu2016,
  title = {Design, Fabrication, and Measurement of Reflective Metasurface for Orbital Angular Momentum Vortex Wave in Radio Frequency Domain},
  author = {Yu, Shixing and Li, Long and Shi, Guangming and Zhu, Cheng and Zhou, Xiaoxiao and Shi, Yan},
  date = {2016-03-21},
  journaltitle = {Applied Physics Letters},
  shortjournal = {Appl. Phys. Lett.},
  volume = {108},
  number = {12},
  pages = {121903},
  issn = {0003-6951, 1077-3118},
  doi = {10.1063/1.4944789},
  url = {https://pubs.aip.org/apl/article/108/12/121903/29945/Design-fabrication-and-measurement-of-reflective},
  urldate = {2026-08-28},
  langid = {english},
}

@article{yue2016,
  title = {Vector {{Vortex Beam Generation}} with a {{Single Plasmonic Metasurface}}},
  author = {Yue, Fuyong and Wen, Dandan and Xin, Jingtao and Gerardot, Brian D. and Li, Jensen and Chen, Xianzhong},
  date = {2016-09-21},
  journaltitle = {ACS Photonics},
  volume = {3},
  number = {9},
  pages = {1558--1563},
  publisher = {American Chemical Society},
  issn = {2330-4022, 2330-4022},
  doi = {10.1021/acsphotonics.6b00392},
  url = {https://pubs.acs.org/doi/10.1021/acsphotonics.6b00392},
  urldate = {2026-03-27},
  langid = {english},
}

@article{yue2017,
  title = {Multichannel {{Polarization}}-{{Controllable Superpositions}} of {{Orbital Angular Momentum States}}},
  author = {Yue, Fuyong and Wen, Dandan and Zhang, Chunmei and Gerardot, Brian D. and Wang, Wei and Zhang, Shuang and Chen, Xianzhong},
  date = {2017-04},
  journaltitle = {Advanced Materials},
  shortjournal = {Adv. Mater.},
  volume = {29},
  number = {15},
  pages = {1603838},
  issn = {0935-9648, 1521-4095},
  doi = {10.1002/adma.201603838},
  url = {https://advanced.onlinelibrary.wiley.com/doi/10.1002/adma.201603838},
  urldate = {2026-03-27},
  langid = {english},
}

@article{zhang2019b,
  title = {Plasmonic Metasurfaces with 42.3\% Transmission Efficiency in the Visible},
  author = {Zhang, Jihua and ElKabbash, Mohamed and Wei, Ran and Singh, Subhash C. and Lam, Billy and Guo, Chunlei},
  date = {2019-06-12},
  journaltitle = {Light: Science \& Applications},
  shortjournal = {Light Sci. Appl.},
  volume = {8},
  number = {1},
  pages = {53},
  issn = {2047-7538},
  doi = {10.1038/s41377-019-0164-8},
  url = {https://www.nature.com/articles/s41377-019-0164-8},
  urldate = {2026-04-29},
  langid = {english},
}

@article{zhao2011,
  title = {Manipulating Light Polarization with Ultrathin Plasmonic Metasurfaces},
  author = {Zhao, Yang and Al\`u, Andrea},
  date = {2011-11-16},
  journaltitle = {Physical Review B},
  shortjournal = {Phys. Rev. B},
  volume = {84},
  number = {20},
  pages = {205428},
  publisher = {American Physical Society},
  issn = {1098-0121, 1550-235X},
  doi = {10.1103/PhysRevB.84.205428},
  url = {https://link.aps.org/doi/10.1103/PhysRevB.84.205428},
  urldate = {2026-03-27},
  langid = {english},
}

@article{zhao2013,
  title = {Tailoring the {{Dispersion}} of {{Plasmonic Nanorods To Realize Broadband Optical Meta-Waveplates}}},
  author = {Zhao, Yang and Al\`u, Andrea},
  date = {2013-03-13},
  journaltitle = {Nano Letters},
  shortjournal = {Nano Lett.},
  volume = {13},
  number = {3},
  pages = {1086--1091},
  issn = {1530-6984, 1530-6992},
  doi = {10.1021/nl304392b},
  url = {https://pubs.acs.org/doi/10.1021/nl304392b},
  urldate = {2023-08-03},
  langid = {english},
}

@article{zheng2015a,
  title = {Metasurface Holograms Reaching 80\% Efficiency},
  author = {Zheng, Guoxing and M\"uhlenbernd, Holger and Kenney, Mitchell and Li, Guixin and Zentgraf, Thomas and Zhang, Shuang},
  date = {2015-04},
  journaltitle = {Nature Nanotechnology},
  shortjournal = {Nat. Nanotechnol.},
  volume = {10},
  number = {4},
  pages = {308--312},
  issn = {1748-3387, 1748-3395},
  doi = {10.1038/nnano.2015.2},
  url = {https://www.nature.com/articles/nnano.2015.2},
  urldate = {2023-08-03},
  langid = {english},
}

@software{Tidy3d,
  title = {{{Tidy3D}}: {{Electromagnetic Solver}} for {{FDTD}}},
  author = {{Flexcompute Inc.}},
  date = {2025},
  url = {https://www.flexcompute.com/tidy3d},
  urldate = {2026-10-15},
  version = {2.9.1},
}

@article{meep,
title = {Meep: A flexible free-software package for electromagnetic simulations by the FDTD method},
journal = {Computer Physics Communications},
volume = {181},
number = {3},
pages = {687-702},
year = {2010},
issn = {0010-4655},
doi = {https://doi.org/10.1016/j.cpc.2009.11.008},
url = {https://www.sciencedirect.com/science/article/pii/S001046550900383X},
author = {Ardavan F. Oskooi and David Roundy and Mihai Ibanescu and Peter Bermel and J.D. Joannopoulos and Steven G. Johnson},
}

\end{document}